\documentclass[11pt]{article}
\usepackage{amsmath,amssymb}
\newcommand{\ft}[2]{{\textstyle\frac{#1}{#2}}}

\newsavebox{\uuunit}
\sbox{\uuunit}
    {\setlength{\unitlength}{0.825em}
     \begin{picture}(0.6,0.7)
        \thinlines
        \put(0,0){\line(1,0){0.5}}
        \put(0.15,0){\line(0,1){0.7}}
        \put(0.35,0){\line(0,1){0.8}}
       \multiput(0.3,0.8)(-0.04,-0.02){10}{\rule{0.5pt}{0.5pt}}
     \end {picture}}

\numberwithin{equation}{section}
\begin{document}
\begin{titlepage}
\begin{center}
\hfill LMU-ASC 45/08 \\
\hfill ITP-UU-08/47 \\
\hfill SPIN-08/37  \\
\vskip 6mm

%%%%%%%%%%%%%%%%%%%%%%%%%%%%%%%%%%%%%%%%%%
{\Large \textbf{Subleading and non-holomorphic corrections to\\[2mm] 
N=2 BPS black hole entropy }}
%%%%%%%%%%%%%%%%%%%%%%%%%%%%%%%%%%%%%%%%%%%
\vskip 8mm

\textbf{G. L.~Cardoso$^{a}$, B. de Wit$^b$ and S.~Mahapatra$^{c}$}

\vskip 4mm
$^a${\em Arnold Sommerfeld Center for Theoretical Physics\\
Department f\"ur Physik,
Ludwig-Maximilians-Universit\"at M\"unchen, Munich, Germany}\\
{\tt gabriel.cardoso@physik.uni-muenchen.de}\\[1mm]
$^b${\em Institute for Theoretical Physics} and {\em Spinoza
  Institute,\\ Utrecht University, Utrecht, The Netherlands}\\
{\tt  B.deWit@uu.nl} \\[1mm]
$^c${\em 
Physics Department, Utkal University, 
Bhubaneswar 751 004, India}\\
{\tt swapna@iopb.res.in}\\[1mm]

\end{center}
\vskip .2in
%%%%%%%%%%%%%%%%%%%%%%%%%%%%%%%%%%%%%%%%%%%%%%%%%%%%%
\begin{center} {\bf ABSTRACT } \end{center}
\begin{quotation}\noindent
  BPS black hole degeneracies can be expressed in terms of an inverse
  Laplace transform of a partition function based on a mixed
  electric/magnetic ensemble, which involves a non-trivial integration
  measure. This measure has been evaluated for black holes with
  various degrees of supersymmetry and for N=4 supersymmetric black
  holes all results agree. It generally receives contributions from
  non-holomorphic corrections. An explicit evaluation of these
  corrections in the context of the effective action of the FHSV model
  reveals that these are related to, but quantitatively different
  from, the non-holomorphic corrections to the topological string,
  indicating that the relation between the twisted partition functions
  of the latter and the effective action is more subtle than has so
  far been envisaged. The effective action result leads to a duality
  invariant BPS free energy and arguments are presented for the
  existence of consistent non-holomorphic deformations of special
  geometry that can account for these effects.  A prediction is given
  for the measure based on semiclassical arguments for a class of N=2
  black holes. Furthermore an attempt is made to confront some of the
  results of this paper with a recent proposal for the microstate
  degeneracies of the STU model.
\end{quotation}

\vfill
%%%%%%%%%%%%%%%%%%
%\flushleft{\today}
%%%%%%%%%%%%%%%%%%
\end{titlepage}
%%%%%%%%%%%%%%%%
\eject
%%%%%%%%%%%%%%%%
%%%%%%%%%%%%%%%%%%%%%%%%%%%%%%%%%%%%%%%%%%%%%%%%%%%%%%%%%%%%%%%%%%%%%%
%%%%%%%%%%%%%%%%%%%%%%%%%%%%%%%%%%%%%%%%%%%%%%%%%%%%%%%%%%%%%%%%%%%%%%
\section{Introduction}
\label{sec:introduction}
\setcounter{equation}{0}
%%%%%%%%%%%%%%%%%%%%%%%%%%%%%%%%%%%%%%%%%%%%%%%%%%%%%%%%%%%%%%%%%%%%%%
The degeneracy of BPS states of certain wrapped brane/string
configurations defines a microscopic entropy which, in quite a number
of cases, has been successfully compared to the macroscopic entropy of
supersymmetric black hole solutions in the corresponding effective
supergravity theories. Agreement is usually obtained in the limit
where charges are large \cite{Strominger:1996sh}, because in that
limit one can make use of the Cardy formula for the underlying
conformal field theory. The macroscopic entropy is not necessarily
identified with a quarter of the horizon area, since there are
corrections associated with higher-derivative couplings
\cite{Maldacena:1997de,Vafa:1997gr,LopesCardoso:1998wt}.  More
recently, it was proposed that the entropy of four-dimensional
BPS black holes with
$N=2$ supersymmetry is related to a partition function based on a
mixed ensemble defined in terms of magnetic charges and electrostatic
potentials. Discarding non-holomorphic corrections this partition
function equals the modulus square of the topological string partition
function \cite{Ooguri:2004zv}. On the basis of this relation it was
concluded that the microscopic black hole degeneracies can be
retrieved from the topological string partition function by an inverse
Laplace transform. This observation gave new impetus to studying the
relation between microscopic and macroscopic descriptions of black
holes on the one hand, and the relation between black hole
degeneracies and the topological string on the other. (See, for
instance, \cite{Dabholkar:2004yr,Ooguri:2005vr,
  Dabholkar:2005dt,Shih:2005he,Pestun:2005ni,LopesCardoso:2006bg,
  Gaiotto:2006ns,Beasly:2006us,deBoer:2006vg, Denef:2007vg}.)

As was readily understood the proper definition of the inverse Laplace
integral is subtle for reasons of convergence and in view of
ambiguities in choosing the integration contours. The issue of
non-holomorphicity did not enter into the original proposal. Early
discussions can be found in
\cite{LopesCardoso:1999ur,Verlinde:2004ck,Sen:2005pu}. Non-holomorphic
terms are essential for duality invariance, and indeed such terms were
encountered when confronting the asymptotic results from microstate
counting with macroscopic results based on effective actions
\cite{Cardoso:2004xf,Jatkar:2005bh,LopesCardoso:2006bg}. They involve
terms originating from higher-order interactions that contain the
square of the Riemann tensor, such as the ones that were determined in
\cite{Harvey:1996ir,Harvey:1996ts,Gregori:1997hi}, which are part of
the effective field theory. The presence of non-holomorphic
corrections can also be inferred from the relation with the
topological string, where they are encoded in the so-called
holomorphic anomaly equations \cite{Bershadsky:1993cx}.

At an early stage there were strong indications that the inverse
Laplace transform must involve a non-trivial integration measure
(which will contribute to the subleading entropy corrections for large
black holes in the limit of large charges), so that (subleading)
non-holomorphic corrections can always be factored out from the mixed
partition function and absorbed into this measure.  Therefore a
further understanding of these matters will ultimately depend on how
well the measure factor can be understood. A strong argument in favour
of the measure was based on the invariance under duality, as the
partition function for the mixed ensemble does not transform simply
under electric/magnetic duality. An alternative starting point
\cite{deWit-Toronto,LopesCardoso:2006bg} can be based on an ensemble
of electric and magnetic charges, which is manifestly invariant under
duality.  From this set-up the previous formulation based on the mixed
partition function can be reobtained in the semiclassical
approximation, but, as it turns out, it is now accompanied by a
non-trivial measure factor. Independently, a direct evaluation of the
mixed partition function from specific microscopic degeneracy formulae
also revealed the presence of a measure factor \cite{Shih:2005he}, and
it was shown that for large charges these measure factors were in fact
equal \cite{LopesCardoso:2006bg,de Wit:2007dn}.

Somewhat unfortunately, the examples studied in
\cite{Cardoso:2004xf,Shih:2005he,Jatkar:2005bh,LopesCardoso:2006bg}
did not pertain to genuine $N=2$ supersymmetric string
compactifications (the work reported in
\cite{Maldacena:1997de,Vafa:1997gr,LopesCardoso:1998wt} is an
exception to this), but to compactifications with $N=4$ supersymmetry.
The latter were then treated in the context of an $N=2$ supersymmetric
truncation with minor modification such as to account for the four
extra graviphotons (leading to eight extra charges) and moduli
provided by the two additional gravitino
supermultiplets.\footnote{%%%%%%%%%%%%%%%%%%%%%%%%%%%%%%%%%%%%%%%%%%
  This is in contrast with the work on small black holes reported in
  \cite{Dabholkar:2005by}. }
%%%%%%%%%%%%%%%%%%%%%%%%%%%%%%%%%%%%%%%%%%%%%%%%%%%%%%%%%%%%%%%%%%
The purpose of the present paper is to study applications that pertain
to genuine $N=2$ supersymmetric models in four space-time dimensions,
where such modifications are unnecessary. The problem with generic
$N=2$ supersymmetric compactifications is, however, that there are not
many cases where it is possible to make direct comparisons with
microstate counting and, at the same time, exact duality invariance is
rather rare. There are a few models which stand out in this respect,
such as the FHSV model \cite{Ferrara:1995yx} and the STU model
\cite{Sen:1995ff,Gregori:1999ns}, which exhibit both exact S- and
T-dualities and for which microstate degeneracy formulae have recently
been proposed \cite{David:2007tq}. For $N=2$ models based on compact
Calabi-Yau spaces, the measure factor has recently been evaluated at
strong topological string coupling \cite{Denef:2007vg}. We will show
that this result disagrees with the semiclassical prediction relevant
at weak coupling.  We will comment on this at the end of section
\ref{sec:measure-factor-Omega-1}, where we also compare to results for
the measure factor in $N=4,8$ models.

Special attention is devoted to the issue of non-holomorphic
corrections, which contribute to the measure factor. As it turns out,
the existence of a semiclassical free energy for BPS black holes
(which plays an important role in the variational principle for the
attractor equations) indicates that these corrections must be encoded
in a single real homogeneous function. For $N=4$ black holes this form
of the free energy has been used successfully
\cite{Cardoso:2004xf,LopesCardoso:2006bg}, but in that case the
non-holomorphic corrections are severely restricted, so that the
consequences of this approach were rather minor. Therefore we further
investigate the consequences of this approach in the context of the
FHSV model by concentrating on the requirements posed by the exact
dualities of this model. As an example we derive the subleading
corrections to the function that encodes the effective action and
explicitly compare the result to the topological string for the
genus-1 and genus-2 contributions. As it turns out the results are
clearly different.

Hence the precise relationship between the non-holomorphic terms in
the effective action and those in the topological string partition
functions is not entirely clear. In fact we will present further
evidence that the relation between the functions that encode the
effective action and the partition functions of the topological string
is more subtle than has previously been envisaged. In
\cite{Antoniadis:1993ze,Bershadsky:1993cx} it was shown that certain
string amplitudes are related to the twisted partition functions of
the topological string. These results, however, do not necessarily
imply that the effective action should also have such a direct
relationship, in view of the fact that the effective action
encompasses only the one-particle irreducible diagrams and not the
connected diagrams. As is well known the relation between these two
sets of diagrams proceeds through a Legendre transform. Interestingly
enough, a Legendre transform is also involved when one wishes to
realize the duality transformations in a manifest way in a
field-theoretic context. Here it is important to realize that the
action is not manifestly invariant under symmetries that are induced
by electric/magnetic duality \cite{Gaillard:1981rj,deWit:2001pz}. In
order to obtain manifestly invariant quantities, one may, for
instance, apply a Legendre transform and consider the Hamiltonian
instead. However, in the context of special geometry it is suggestive
to consider the Legendre transform that leads from complex to real
special geometry.  In that case one obtains the so-called Hesse
potential, which is related to the black hole free energy and which is
manifestly duality invariant (this was discussed in
\cite{LopesCardoso:2006bg}). The above scenario for explaining the
discrepancy is admittedly a bit speculative and it is beyond the scope
of this paper to try and work this out further. Obviously, this aspect
has a bearing on the original conjecture of \cite{Ooguri:2004zv}.

Returning to the black holes, there are two aspects that have come
under intense scrutiny lately which will not enter into our analysis.
The first aspect concerns the dependence of the microstate
degeneracies on the asymptotic values of the scalar moduli, i.e., on
the values of the scalar fields at spatial infinity (see, for
instance,
\cite{Sen:2007vb,Denef:2007vg,Dabholkar:2007vk,Sen:2007pg,Cheng:2007ch}).
This dependence is associated with the appearance or disappearance of
multicentered black hole configurations
\cite{Denef:2000nb,Denef:2001xn} for a given total charge.  The second
aspect concerns the so-called entropy enigma, a surprising phenomenon
that may arise at weak topological string coupling
\cite{Denef:2007vg}.  It is based on the fact that there exist
multicentered black hole solutions that carry an entropy that is
vastly larger than the entropy of singlecentered solutions carrying the
same charges.  The occurence of this phenomenon would imply a
breakdown of the conjecture of \cite{Ooguri:2004zv}, which was
supposed to work at weak coupling. It would be difficult to reconcile
this with the fact that the predictions for large black holes have
always been in agreement with semiclassical reasoning. Evidence
against such a breakdown has recently been given in
\cite{Huang:2007sb}.  The approach followed in this paper will not
take into account the two aspects just described and we will assume
that semiclassical arguments do make sense.

%%%%%%%%%%%%%%%%%%%%%%%%%%%%%%%%%%%%%%%%%%%%%%%%%%%%%%%%%%%%%%%%%%%%%
This paper is organized as follows.  Section
\ref{sec:partition-entropy-function} contains a brief review of the
derivation of the measure factor from a duality invariant perspective.
Subsequently the non-holomorphic corrections are incorporated in the
black hole free energy and we discuss the semiclassical approximation.
Section \ref{sec:omega-duality-constraints} describes the consequences
of S- and T-duality invariance for a class of models that contain in
particular the FSHV and the STU models. In section
\ref{sec:measure-factor-Omega-1} the measure factors for the mixed
partition function are evaluated for these models in the semiclassical
approximation. In section \ref{sec:non-holom-corr} non-holomophic
corrections are studied for the FHSV model and compared to the results
for the topological string. Subsequently non-holomorphic deformations
of special geometry are discussed. Section \ref{sec:STU} deals with
the STU model and describes an attempt to reconcile the macroscopic
and microscopic results for the BPS black hole entropy in that model.
Section \ref{sec:conclusion} presents our conclusions.

%%%%%%%%%%%%%%%%%%%%%%%%%%%%%%%%%%%%%%%%%%%%%%%%%%%%%%%%%%%%%%%%%%%%%%
\section{The BPS black hole free energy and the partition function}
\label{sec:partition-entropy-function}
\setcounter{equation}{0}
%%%%%%%%%%%%%%%%%%%%%%%%%%%%%%%%%%%%%%%%%%%%%%%%%%%%%%%%%%%%%%%%%%%%%%
At the field-theoretic level it is known that the attractor equations
that determine the values of the moduli at the black hole horizon
\cite{Ferrara:1995ih,Strominger:1996kf,Ferrara:1996dd}, follow from a
variational principle. This variational principle is described in
terms of a so-called entropy function. There exists an entropy
function for extremal black holes \cite{Sen:2005wa,Sahoo:2006rp},
where the attractor mechanism is induced by the restricted space-time
geometry of the horizon, and one for BPS black holes
\cite{LopesCardoso:2006bg}, where the attractor mechanism follows from
supersymmetry enhancement at the horizon. For $N=2$ supergravity the
relation between these entropy functions has been clarified in
\cite{Cardoso:2006xz}.  To preserve the variational principle when
non-holomorphic corrections are present, it follows that these
corrections must enter into the BPS free energy in a well-defined way.
Requiring the existence of a free energy seems desirable from the
point of view of semiclassical arguments and the relation with black
hole thermodynamics, and it should be interesting to derive this
result directly from an effective action. However, no effective $N=2$
supersymmetric action is known to date that incorporates the
non-holomorphic terms, although partial results are known for $N=1$
\cite{Dixon:1990pc} and from the string amplitudes that are related to
the topological string \cite{Antoniadis:1993ze}. We will discuss this
last relationship in section \ref{sec:non-holom-corr}. At any rate,
the results of this paper indicate that, indeed, one can safely
proceed by checking the internal consistency at the level of the
entropy function, guided by the (partially established) relation with
the full effective action.  This is the underlying strategy of this
paper.

In the first subsection we discuss the definition of the free energy,
and its relation with the black hole partition function and the BPS
entropy function, for a given set of degeneracies and a corresponding
locally supersymmetric effective action. The second subsection
describes the non-holomorphic contributions to the free energy, and
the third subsection deals with the semiclassical approximation.

%%%%%%%%%%%%%%%%%%%%%%%%%%%%%%%%%%%%%%%%%%%%%%%%%%%%%%%%%%
\subsection{BPS free energy and partition functions}
\label{sec:bps-free-energy}
%%%%%%%%%%%%%%%%%%%%%%%%%%%%%%%%%%%%%%%%%%%%%%%%%%%%%%%%%%
We consider charged black holes in the context of $N=2$ supergravity
in four space-time dimensions, which contains $n+1$ abelian vector
gauge fields, labeled by indices $I,J=0,1,\ldots,n$, so that black
hole solutions can carry $2(n+1)$ possible electric and magnetic
charges. The theory describes the supergravity fields and $n$ vector
multiplets (the extra index $I=0$ accounts for the gauge field that
belongs to the supergravity multiplet), and possibly a number of
hypermultiplets which will only play an ancillary role. A partition
sum over a canonical ensemble of corresponding BPS black hole
microstates is defined as follows,
\begin{equation}
  \label{eq:partition}
  Z(\phi,\chi) = \sum_{\{p,q\}} \;   d(p,q)
  \, \mathrm{e}^{\pi [q_I \phi^I - p^I \chi_I]} \,,
\end{equation}
where $d(p,q)$ denotes the degeneracy of the black hole microstates
with given magnetic and electric charges equal to $p^I$ and $q_I$,
respectively. This expression is consistent with electric/magnetic
duality, provided that the electro- and magnetostatic potentials
$(\phi^I,\chi_I)$ transform as a symplectic vector, just as the
charges $(p^I,q_I)$, while the degeneracies $d(p,q)$ transform as
functions of the charges under the duality. In case that the duality
is realized as a symmetry, then the $d(p,q)$ should be invariant.

Viewing $Z(\phi,\chi)$ as an analytic function in $\phi^I$ and
$\chi_I$, the degeneracies $d(p,q)$ can be retrieved by an inverse
Laplace transform,
\begin{equation}
  \label{eq:inverse-LT}
  d(p,q) \propto \int \;{\mathrm{d}\phi^I\,\mathrm{d}\chi_I}\;
  Z(\phi,\chi) \; \mathrm{e}^{\pi [-q_I \phi^I + p^I \chi_I]} \,,
\end{equation}
where the integration contours run, for instance, over the intervals
$(\phi-\mathrm{i}, \phi +\mathrm{i})$ and $(\chi-\mathrm{i}, \chi
+\mathrm{i})$ (we are assuming an integer-valued charge lattice).
Obviously, this makes sense as long as $Z(\phi,\chi)$ is formally
periodic under shifts of $\phi$ and $\chi$ by multiples of
$2\mathrm{i}$.

Identifying the logarithm of $Z(\phi,\chi)$ with a free energy, it is
expected that this expression has a field-theoretic counterpart,
because the electrostatic and magnetostatic fields appear as some of
the scalar moduli in the field-theoretic description. Indeed, such a
free energy function exists and it is contained in the so-called BPS
entropy function. Stationary points of this entropy function are
subject to the attractor equations which fix the value of the moduli
at the black hole horizon, and the value of the entropy function at
the stationary point equals the macroscopic entropy. The latter is a
function of the charges and it equals the Legendre transform of the
free energy.  The BPS entropy function was originally proposed in
\cite{Behrndt:1996jn} for actions that are at most quadratic in
space-time derivatives and its generalization to higher derivatives
was discussed in \cite{LopesCardoso:2006bg}. It is natural to identify
the partition function \eqref{eq:partition} with the exponent of the
relevant free energy, which is contained in the entropy function. In
the case at hand, where one considers functions of real potentials
$(\phi^I,\chi_I)$, this free energy equals twice the so-called Hesse
potential $\mathcal{H}$, which depends on the holomorphic function
that encodes the $N=2$ supergravity theory of the vector multiplet
sector \cite{de Wit:1984pk}. In the notation of
\cite{LopesCardoso:2006bg}, we write
\begin{equation}
  \label{eq:partition-hesse}
   \sum_{\{p,q\}} \; d(p,q) 
  \,\mathrm{e}^{\pi[q_I \phi^I - p^I \chi_I]} \sim 
  \sum_{\rm shifts} \;
  \mathrm{e}^{2\pi\,\mathcal{H}(\phi/2,\chi/2)}
  \,.
\end{equation}
The Hesse potential is a macroscopic quantity which does not in
general exhibit the periodicity that is characteristic for the
partition function.  Therefore, the right-hand side of
(\ref{eq:partition-hesse}) requires an explicit periodicity sum over
discrete imaginary shifts of the $\phi^I$ and
$\chi_I$.\footnote{%%%%%%%%%%%%%%%%%%%%%%%%%%%%%%%%%%%%%%%%%%%%%%%%
  In case that the Hesse potential exhibits a periodicity with a
  multiple of the periodicity interval, then the sum over the
  imaginary shifts will have to be modded out appropriately such as to
  avoid overcounting. } %%%%%%%%%%%%%%%%%%%%%%%%%%%%%%%%%%%%%%%%%%%%%%
In the inverse Laplace integral \eqref{eq:inverse-LT} we expect that
this periodicity sum can be incorporated into the integration contour.

It is in general difficult to find an explicit representation for the
Hesse potential. The standard way to encode the effective supergravity
theory (as far as the vector multiplet sector is concerned), is in
terms of a holomorphic function of the complex scalar fields $Y^I$,
and the resulting geometric structure is known as special geometry.
Here one identifies a symplectic vector by combining the scalars $Y^I$
with the holomorphic derivatives $F_I$ of the function $F(Y)$, which
transforms under duality precisely as the charges $(p^I,q_I)$.  Of
course, this leaves several options for parametrizing the models, and
the obvious one that leaves the symplectic structure intact is to
choose real variables equal to the electro- and magnetostatic
potentials \cite{LopesCardoso:2000qm},
\begin{equation}
  \label{eq:real-special-vars}
  \phi^I = Y^I + \bar Y^I\,,\qquad \chi_I = F_I + \bar F_{\bar I}\,. 
\end{equation}
In these variables one obtains the Hesse potential as a Legendre
transform of the imaginary part of $F(Y)$ with respect to the
imaginary part of the $Y^I$. This is precisely equal to one-half of
the free energy $\mathcal{F}(Y,\bar{Y})$, defined in complex
coordinates, that we will discuss momentarily. Substitution of these
relations leads to,
\begin{equation}
  \label{eq:partition1}
   \sum_{\{p,q\}} \;   d(p,q)
   \, \mathrm{e}^{\pi [q_I (Y^I+\bar Y^I) - p^I (F_I +\bar{F}_I)]} \sim 
   \sum_{\rm shifts} \;
  \mathrm{e}^{\pi \,\mathcal{F} (Y,\bar Y) }
   \,,
\end{equation}
but now the definition of the shifts has become very subtle as they
still refer to imaginary values of $\phi^I$ and $\chi_I$. This
subtlety should again be reflected in the choice of the integration
contours in the inverse Laplace transform. We emphasize that at this
point we are assuming that $F(Y)$ is a holomorphic function which is
homogeneous of second degree, although so far we did not make use of
this. The equation \eqref{eq:partition1} is the conjectured relation
between the microscopic data, defined in terms of the degeneracies
$d(p,q)$, and the field-theoretic data, encoded in the free energy
$\mathcal{F}$. In this section we will derive the expression for this
free energy in terms of derivatives of the function $F$ in the
presence of subleading and non-holomorphic corrections, and discuss
some consequences of this result. The expression for the free energy
follows from the requirement that the attractor equations are based on
a variational principle. The reason for adopting this procedure is
that in the presence of non-holomorphic corrections, the effective
action is not fully known and hence cannot be used directly to define
the free energy. We already discussed this strategy at the beginning
of this section.

Postponing the discussion of various subtleties and generalizations,
we consider a variable change from the real variables
$(\chi^I,\phi_I)$ to the complex variables $Y^I$ in the integral
\eqref{eq:inverse-LT}, replacing $Z(\phi,\chi)$ by $\exp[2\pi
\,\mathcal{H}(\phi/2,\chi/2)]$, and subsequently by
$\exp[\pi\,\mathcal{F}(Y,\bar Y)]$ when changing variables. This leads
to the integral,
\begin{equation}
  \label{eq:laplace1}
  \begin{split}
    d(p,q) &\propto \int \; \mathrm{d} (Y+\bar Y)^I\; \mathrm{d}
    (F+{\bar F})_I
    \;\mathrm{e}^{\pi\,\Sigma(Y,\bar Y,p,q)} \\
    &\propto \int \; \mathrm{d} Y^I\, \mathrm{d}\bar
    Y^I\;\Delta(Y,\bar Y)\; \mathrm{e}^{\pi\,\Sigma(Y,\bar
    Y,p,q)}\;, 
  \end{split} 
\end{equation}
where $\Delta(Y,\bar Y)$ denotes the Jacobian associated with the
change of integration variables $(\phi,\chi)\to (Y,\bar{Y})$,
\begin{equation}
  \label{eq:simple-measure}
    \Delta(Y,\bar Y) = 
    \left\vert\det[\,{\rm Im}\,2\,F_{KL} ]\right\vert \,, 
\end{equation}
and $\Sigma$ denotes the BPS entropy function which decomposes
according to
\begin{equation}
  \label{eq:Sigma-simple}
  \Sigma(Y,\bar Y,p,q) =  \mathcal{F}(Y,\bar Y)
  - q_I   (Y^I+\bar Y^I)   + p^I (F_I+\bar F_I)  \;. 
\end{equation}
Here $p^I$ and $q_I$ couple to the corresponding magneto- and
electrostatic potentials (c.f. \eqref{eq:real-special-vars}) at the
horizon in a way that is consistent with electric/magnetic duality.
Furthermore, $\mathcal{F}(Y,\bar Y)$ represents the free energy
alluded to earlier. In the following we will consider its definition.

The free energy $\mathcal{F}$ has the property that its
variations take the form, 
\begin{equation}
  \label{eq:delta-mathcal-F}
  \delta\mathcal{F} = \mathrm{i} ( {Y}^I - {\bar Y}^{I})\, \delta 
  ( F_{I} + \bar F_I)  - \mathrm{i} ( F_I - {\bar F}_{{I}})\, 
  \delta(Y^I+ \bar Y^{I})  \;,
\end{equation}
so that the variation of the entropy function $\Sigma$ with respect to the
$Y^I$, while keeping the charges fixed, yields the black hole
attractor equations, 
\begin{equation}
  \label{eq:attractor}
  Y^I-\bar Y^I = \mathrm{i} p^I\,,\qquad
  F_I(Y) - \bar F_{I}(\bar Y) = \mathrm{i} q_I\,. 
\end{equation}
These equations determine the values of the $Y^I$ at the black hole
horizon in terms of the charges. Under the mild assumption that the
matrix $N_{IJ} = 2\,\mathrm{Im}\, F_{IJ}$ is non-degenerate, it thus
follows that stationary points of $\Sigma$ must satisfy the attractor
equations.

One can now evaluate the integral \eqref{eq:laplace1} in the
semiclassical approximation and show that the answer takes the form,
\begin{equation}
  \label{eq:complex-saddle-holo}
    d(p,q) =  \mathrm{e}^{{\cal S}_{\rm macro}(p,q)} \;,
\end{equation}
where $\mathcal{S}_\mathrm{macro}(p,q)$ equals the value of
$\pi\,\Sigma$ taken at the saddle point. This is a gratifying result
as we correctly recover the classical result, provided a free energy
function exists with the required properties. In principle, we should
have included the measure factor \eqref{eq:simple-measure} when
expanding around the saddle point but these contributions are
suppressed in the limit of large charges, where all the charges and
the fields $Y^I$ are scaled uniformly. 

Before continuing and discussing the free energy in further detail, we
wish to emphasize that the scalar fields belonging to the vector
multiplets are projectively defined in the underlying superconformal
framework used for constructing the effective supergravity theory. On
the other hand, the fields $Y^I$ must have been given an intrinsic
normalization as follows from the observation that both sides of the
attractor equations scale differently in view of the fact that the
charges are constant. This is also obvious from the equation $q_IY^I
-p^IF_I= -\mathrm{i}(\bar Y^IF_I- Y^I\bar F_I)$, which holds generally
at the attractor point. Indeed we have adopted a normalization
condition on the $Y^I$ such that they are no longer
subject to these projective redefinitions.\footnote{%%%%%%%%%%%%%%%%%%
  To be specific, the original (projectively defined) fields $X^I$ and
  the normalized fields $Y^I$ are related by \cite{Behrndt:1996jn},
  \begin{equation}
    \label{eq:Y-X-horizon}
    Y^I = \frac{\bar Z\,X^I}{\sqrt{\mathrm{i}\,(\bar X^IF_I(X)-\bar
    F_I(\bar X) X^I)}} \,, 
  \end{equation}
  where 
  \begin{eqnarray}
    \label{eq:Y-Z}
    Z = \frac{p^I F_I(X) - q_I X^I}{\sqrt{\mathrm{i}\,(\bar X^IF_I(X)-\bar
    F_I(\bar X) X^I)}}  \,.
\end{eqnarray}
This latter quantity is sometimes referred to as the holomorphic BPS
mass. Note that the $Y^I$ are invariant under uniform complex
rescalings of the underlying variables $X^I$. \label{footnote-XY} } % 
%%%%%%%%%%%%%%%%%%%%%%%%%%%%%%%%%%%%%%%%%%%%%%%%%%%%%%%%%%%%%  
In the case that the function $F(Y)$ is holomorphic and homogenous of
second degree, the expression for the free energy is known and equal
to $\mathcal{F}(Y,\bar Y)= -\mathrm{i}(\bar Y^IF_I- Y^I\bar F_I)$.
Indeed this expression satisfies \eqref{eq:delta-mathcal-F} by virtue
of the homogeneity of the function $F(Y)$ \cite{Behrndt:1996jn}.

However, in reality, the function $F(Y)$ will depend also on an extra
complex field $\Upsilon$ which is equal to the lowest-dimensional
component of the square of the Weyl multiplet. The presence of this
field encodes interactions in the effective field theory proportional
to the square of the Weyl tensor. Supersymmetry requires the function
$F(Y,\Upsilon)$ to remain holomorphic and homogeneous of second
degree,
\begin{equation}
  \label{eq:homogeneous-F}
  F(\lambda Y,\lambda^2 \Upsilon) = \lambda^2\, F( Y,\Upsilon)\,.
\end{equation}
The BPS free energy takes the following form in the presence of
$\Upsilon$-dependent terms,
\begin{equation}
  \label{eq:free-energy-phase}
  \mathcal{F}(Y,\bar Y,\Upsilon,\bar\Upsilon)= - \mathrm{i} \left( {\bar Y}^I
  F_I - Y^I {\bar F}_I
  \right) - 2\mathrm{i} \left( \Upsilon F_\Upsilon - \bar \Upsilon
  \bar F_{\Upsilon}\right)\,,
\end{equation}
where $F_\Upsilon= \partial F/\partial\Upsilon$. And again, this free
energy satisfies \eqref{eq:delta-mathcal-F} by virtue of the
(modified) homogeneity property \eqref{eq:homogeneous-F}, where $F(Y)$
and $F_I(Y)$ are everywhere replaced by $F(Y,\Upsilon)$ and
$F_I(Y,\Upsilon)$, and where $\Upsilon$ is kept fixed under the
variation. Note that the definition \eqref{eq:free-energy-phase} is
consistent with electric/magnetic duality
\cite{deWit:1996ix}. Furthermore, an encouraging feature is that the
expression \eqref{eq:free-energy-phase} follows directly when
evaluating the Hesse potential based on the holomorphic function $F$
in the presence of $\Upsilon$-dependent terms, without making any 
reference to the attractor equations \cite{LopesCardoso:2006bg}.

The BPS attractor equations impose a constant real value for
$\Upsilon$, namely $\Upsilon=-64$. This implies that the terms
proportional to positive powers of $\Upsilon$ encode subleading
contributions to the entropy. The reason for this is that the
attractor equations and the entropy function scale uniformly under
simultaneous scale transformations of the $Y^I$ and $\Upsilon$ fields
according to \eqref{eq:homogeneous-F}, provided we scale the charges
accordingly.  The fact that the attractor equations fix $\Upsilon$ to
a constant affects this scaling property.  This phenomenon has been
successfully demonstrated in \cite{LopesCardoso:1998wt}, following
earlier work in \cite{Maldacena:1997de,Vafa:1997gr}.

%%%%%%%%%%%%%%%%%%%%%%%%%%%%%%%%%%%%%%%%%%%%%%%%%%%%%%%%%%%
\subsection{Non-holomorphic corrections}
\label{sec:non-holomorphic}
%%%%%%%%%%%%%%%%%%%%%%%%%%%%%%%%%%%%%%%%%%%%%%%%%%%%%%%%%%%

A more subtle issue concerns the non-holomorphic corrections to the
entropy function. Already at an early stage \cite{LopesCardoso:1999ur}
it was clear that non-holomorphic corrections were required for
manifest S-duality in $N=4$ supersymmetric heterotic string
compactifications, which have dual realizations as type-II string
theory on $K3\times \mathrm{T}^2$, or M-theory on $K3\times
\mathrm{T}^2\times S^1$. Non-holomorphic modifications signal
departures from the Wilsonian action and are caused by integrating out
the massless modes. These modifications are required in order to
preserve the physical symmetries which cannot be fully realized at the
level of the Wilsonian action. An early example of this can be found
in \cite{Dixon:1990pc}, where it was shown that the gauge coupling
constants become moduli dependent with non-holomorphic corrections.
Applying the $N=2$ attractor equations to this particular situation
reveals the need for non-holomorphic modifications
\cite{LopesCardoso:1999ur}. Specifically, requiring the vector
$(Y^I,F_I)$ to transform consistently under S-duality monodromies, an
S-duality invariant entropy was obtained. The results of this analysis
were also in accord with the results for the non-holomorphic terms
found in the corresponding effective action \cite{Harvey:1996ir}.
Subsequently, but much later, it was demonstrated in
\cite{Cardoso:2004xf} how these results emerge from a semiclassical
approximation of the microscopic degeneracy formula for $N=4$ dyons
\cite{Dijkgraaf:1996it,Shih:2005uc,David:2006yn,
  Banerjee:2008pv,Banerjee:2008pu,Dabholkar:2008zy}.  However, as we
already alluded to in section~\ref{sec:introduction}, the $N=4$
supersymmetric models are of limited use for studying the general
situation as their $\Upsilon$-dependence in $F(Y,\Upsilon)$ is
severely restricted.

Nevertheless, there is one question that can be addressed already at
this stage, namely, whether one can still derive the attractor
equations from a variational principle in the presence of the
non-holomorphic corrections and define a closed expression for the BPS
entropy function and the free energy introduced earlier. To
investigate this question let us evaluate the variation of the free
energy $\mathcal{F}$ defined in \eqref{eq:free-energy-phase} minus the
right-hand side of its expected variation \eqref{eq:delta-mathcal-F},
without making any further assumptions on the function $F$,
\begin{eqnarray}
  \label{eq:delta-mathcal-F-integrate}
  &&{}
  \delta\mathcal{F} - \mathrm{i} ( {Y}^I - {\bar Y}^{I})\, \delta 
  ( F_{I} + \bar F_I)  + \mathrm{i} ( F_I - {\bar F}_{\bar{I}})\, 
  \delta(Y^I+ \bar Y^{I}) = \nonumber\\
  &&\qquad\qquad
  -\mathrm{i}\left(2\,\Upsilon \delta F_\Upsilon+ Y^I\,\delta F_I
  -F_I\,\delta Y^I \right)  + \mathrm{h.c.}\,.
\end{eqnarray}
The right-hand side of the above equation should either vanish, or
become proportional to the variation of a new term, which can then be
absorbed into $\mathcal{F}$. Inspection shows that there are two
obvious solutions. When the function $F$ is homogeneous of second
degree and holomorphic, \eqref{eq:homogeneous-F} implies,
\begin{equation}
  \label{eq:homogeneous}
  2\,\Upsilon F_\Upsilon + Y^I F_I = 2\,F\,,
\end{equation}
so that $2\,\Upsilon \,\delta F_\Upsilon + Y^I\,\delta F_I
-F_I\,\delta Y^I=0$. In that case the right-hand side of
\eqref{eq:delta-mathcal-F-integrate} vanishes, confirming the result
quoted earlier for the holomorphic case. Alternatively, we may relax
the holomorphicity requirement and assume that $F$ (or part of $F$) is
not holomorphic but purely imaginary, so that we can write $F=
2\,\mathrm{i}\Omega(Y,\bar Y,\Upsilon,\bar\Upsilon)$ with $\Omega$ a
real homogeneous function of second degree, which therefore satisfies
$2\,\Upsilon \Omega_\Upsilon + 2\,\bar\Upsilon \Omega_{\bar\Upsilon} +
Y^I \Omega_I +\bar Y^{\bar I}\Omega_{\bar I} = 2\,\Omega$. In
that case the right-hand side of \eqref{eq:delta-mathcal-F-integrate}
vanishes as well. Hence we may write,
\begin{equation}
  \label{eq:F-decomposition}
  F = F^{(0)}(Y,\Upsilon) + 2\mathrm{i}\,\Omega(Y,\bar
  Y,\Upsilon,\bar\Upsilon) \,,
\end{equation}
where the attractor equations \eqref{eq:attractor} retain the same
form, irrespective of the presence of the non-holomorphic terms. The
decomposition \eqref{eq:F-decomposition} is not unique. When the
function $\Omega$ is harmonic, i.e. when it can be written as the sum
of a holomorphic and an anti-holomorphic function, then one may absorb
the holomorphic part into the first term. The anti-holomorphic part
will then not contribute as it will vanish under the holomorphic
derivatives which enter the attractor equations and the free energy.
In practice we will require that $F^{(0)}$ is independent of
$\Upsilon$.

We are not aware of any other general solutions. These two solutions
are the ones that have been discussed before and are consistent with
all known cases. The second option seems to take the form of a
consistent non-holomorphic deformation of special geometry, as we
shall further discuss in section \ref{sec:non-holom-corr}.

%%%%%%%%%%%%%%%%%%%%%%%%%%%%%%%%%%%%%%%%%%%%%%%%%%%%%%%%
\subsection{Semiclassical approximation}
\label{sec:semicl-appr}
%%%%%%%%%%%%%%%%%%%%%%%%%%%%%%%%%%%%%%%%%%%%%%%%%%%%%%%%
Having determined the free energy with possible non-holomorphic
deformations we return to the inverse Laplace integral
\eqref{eq:laplace1}. This integral, defined in the first line of
\eqref{eq:laplace1}, is expressed in terms of $\Sigma$ given in
\eqref{eq:Sigma-simple} with the associated free energy given in
\eqref{eq:free-energy-phase} with $\Upsilon=-64$. In the presence of
non-holomorphic corrections, the function $F$ appearing in these
expressions is the non-holomorphic one introduced in
\eqref{eq:F-decomposition}. These non-holomorphic modifications will
also introduce an explicit modification in the integration measure
$\Delta$, as follows,
\begin{equation}
  \label{eq:laplace-nonholo}
  \begin{split}
    d(p,q) &\propto \int \; \mathrm{d} (Y^I+\bar Y^{\bar I})\; \mathrm{d}
    (F_I +{\bar F}_{\bar I})
    \;\mathrm{e}^{\pi\,\Sigma(Y,\bar Y,p,q)} \\
    &\propto \int \; \mathrm{d} Y^I\, \mathrm{d}\bar
    Y^{\bar I} \;\Delta^-(Y,\bar Y)\; \mathrm{e}^{\pi\,\Sigma(Y,\bar
    Y,p,q)}\;, 
  \end{split} 
\end{equation}
where we now introduce two Jacobian factors, $\Delta^\pm(Y,\bar Y)$,
defined by 
\begin{equation}
  \label{eq:-measure-pm}
    \Delta^\pm(Y,\bar Y) = 
    \left\vert\det\big[{\rm Im}[\,2\,F_{KL}\pm 2\,F_{K\bar L}]
    \big]\right\vert \,. 
\end{equation}
Observe that the mixed derivative satisfies,
\begin{equation}
  \label{eq:2}
  F_{I\bar J} = - \bar F_{\bar J I}\,.
\end{equation}
because of the fact that the non-holomorphic terms are characterized
by the real function $\Omega$. Obviously, the mixed derivatives vanish
when the function $\Omega(Y,\bar Y,\Upsilon,\bar\Upsilon)$ is
harmonic. When this is not the case, we must adopt indices $\bar
I,\bar J,\ldots$ to refer specifically to non-holomorphic coordinates
and derivatives.

Subsequently one evaluates the semiclassical Gaussian integral that
emerges when expanding the exponent in the integrand to second order
in $\delta Y^I$ and $\delta \bar Y^I$ about the attractor point. As it
turns out \cite{LopesCardoso:2006bg}, this can be done in two steps,
because at the saddle point the semiclassical determinant factorizes
into two sub-determinants, one associated with the real and another
one with the imaginary values of the $Y^I$. These two sub-determinants
are precisely equal to $\Delta^+$ and $\Delta^-$, respectively,
defined in \eqref{eq:-measure-pm}. Performing the integral only over
the imaginary parts of the $Y^I$ partially cancels the Jacobian factor
in \eqref{eq:laplace-nonholo}, and one is left with the integral,
\begin{equation}
  \label{eq:laplace-osv}
  d(p,q) \propto \int \; \mathrm{d} \phi \; \sqrt{\Delta^-(p,\phi)} \;
\mathrm{e}^{\pi[\mathcal{F}_{\rm E}(p,\phi)- q_I\phi^I]} \;,
\end{equation}
where 
\begin{equation}
  \label{eq:yphi}
Y^I = \tfrac1{2}(\phi^I + \mathrm{i} p^I) \;.
\end{equation}
Hence this result takes the form of the OSV integral
\cite{Ooguri:2004zv}, with an extra integration measure
$\sqrt{\Delta^-}$. In view of the original setting in terms of the
Hesse potential, we expect that the integration contours in
\eqref{eq:laplace-osv} should be taken along the imaginary axes. The
free energy associated with the mixed ensemble, $\mathcal{F}_{\rm
  E}(p,\phi)$, reads as follows, 
\begin{equation}
  \label{eq:osv2-nonholo}
  \mathcal{F}_{\rm E}(p,\phi) = 4\,\Big[ {\rm Im}\,F(Y,\bar
  Y,\Upsilon,\bar \Upsilon) -
  \Omega(Y,\bar Y,\Upsilon,\bar\Upsilon)\Big]_{Y^I=(\phi^I+ \mathrm{i} 
  p^I)/2}  \,.
\end{equation}
The remaining attractor equations read, $q_I= \partial
\mathcal{F}_{\rm E}/\partial \phi^I$. We note the presence of
the term proportional to $\Omega$, which partially cancels the
$\Omega$-dependence in the function $F$. The reader may verify that,
when $\Omega$ is harmonic, everything can be expressed in terms of the
imaginary part of the properly modified holomorphic function $F$.

It remains to complete the semiclassical approximation and perform the
integral over the $\phi^I$. This gives the result,
\begin{equation}
  \label{eq:complex-saddle}
    d(p,q) = \sqrt{
  \left\vert \,\frac{\Delta^-(Y,\bar Y)}{\Delta^+(Y,\bar Y)}\right\vert }_{\rm
   attractor} \; \mathrm{e}^{{\cal S}_{\rm macro}(p,q)} \;.
\end{equation}
In the absence of non-holomorphic corrections the ratio of the two
determinants is equal to unity and one recovers precisely the
macroscopic entropy, as in \eqref{eq:complex-saddle-holo}.

Inverting \eqref{eq:laplace-osv} to a partition sum over a mixed
ensemble, one finds,
\begin{eqnarray}
  \label{eq:partition1-osv} 
    Z(p,\phi) &=&
    \sum_{\{q\}} \; d(p,q) \, \mathrm{e}^{\pi \, q_I \phi^I } \nonumber\\
    &\sim&{} 
  \sum_{\rm shifts} \;
  \sqrt{\Delta^-(p,\phi)}\;
    \mathrm{e}^{\pi\,\mathcal{F}_{\rm E}(p,\phi)}  \;.
\end{eqnarray}
The function $\mathcal{F}_{\mathrm{E}}$ is not
duality invariant and the invariance is only recaptured when
completing the saddle-point approximation with respect to the fields
$\phi^I$. Therefore an evaluation of (\ref{eq:laplace-osv}) beyond the
saddle-point approximation will most likely give rise to a violation
of (some of) the duality symmetries.

To discuss the validity of the semiclassical approximation, we recall
that the entropy function is homogeneous of degree two under uniform
rescalings of the charges, $(p^I,q_I)$, and the fields $Y^I$ and
$\sqrt{\Upsilon}$ and their complex conjugates. However, $\Upsilon$
will take a fixed value as a result of the attractor equations.
Therefore $\Upsilon$-dependent terms affect the uniform scaling and,
under the assumption that only positive powers of $\Upsilon$ appear,
are associated with subleading corrections. The leading terms in the
BPS entropy function scale quadratically, and so does the entropy. On
the other hand, the leading contributions to the determinant factors
scale with zero weight. Hence the latter terms do not have to be expanded
about the saddle point as they would yield contributions with negative
scaling weights. The semiclassical approximation thus pertains to all
terms that scale with non-negative scaling weights. Therefore
subleading corrections to the entropy function with zero weight are
comparable to the {\it leading} terms in the determinant factors.
Assuming that $\Omega$ is at least proportional to $\Upsilon$ or its
complex conjugate, we have to include the terms in $\Omega$ that are
linear in $\Upsilon$, but we can suppress them in the determinants.
In that case the prefactor in \eqref{eq:complex-saddle} equals unity.

Hence we expect that the semiclassical approximation is reliable for
the leading and subleading terms in the entropy. The consistency of
this approach has been verified in many cases, but mainly for large
black holes in $N=4$ supersymmetric string compactifications based on
an $N=2$ supersymmetric description
\cite{LopesCardoso:1999ur,Cardoso:2004xf,Shih:2005he,Jatkar:2005bh,LopesCardoso:2006bg}.
Obviously this result is not compatible with the so-called entropy
enigma, found in \cite{Denef:2007vg}. In the case of small black
holes, where the leading contribution is absent, the above arguments
do not quite apply and the semiclassical approximation breaks down,
although the next-to-leading part in the entropy can still be
calculated reliably
\cite{LopesCardoso:1999ur,Dabholkar:2004yr,Dabholkar:2005by}.

%%%%%%%%%%%%%%%%%%%%%%%%%%%%%%%%%%%%%%%%%%%%%%%%%%%%%%%%%%%%%%%%%%%%%%
\section{Constraints on $\Omega$ due to exact duality symmetries}
\label{sec:omega-duality-constraints}
\setcounter{equation}{0}
%%%%%%%%%%%%%%%%%%%%%%%%%%%%%%%%%%%%%%%%%%%%%%%%%%%%%%%%%%%%%%%%%%%%%%
In this section we consider specific $N=2$ models with exact duality
symmetry groups.  Two such models are the FHSV \cite{Ferrara:1995yx}
and the STU model \cite{Sen:1995ff,Gregori:1999ns}. Their symmetries
constrain the form of the real homogeneous function $\Omega$ in
\eqref{eq:F-decomposition}, because the corresponding monodromies
imply specific transformation rules for the derivatives of $\Omega$.
We begin by discussing exact duality symmetries in the context of a
larger class of models, which will enable us to make contact with
previous work on BPS black hole entropy applied to various string
compactifications invariant under 8 or 16 supersymmetries. The results
of this section will then be used in later sections. 

The FHSV model \cite{Ferrara:1995yx} is a model with 8
supersymmetries.  Its type-II realization corresponds to the
compactification on the Enriques Calabi-Yau three-fold, which is
described as an orbifold
$(\mathrm{T}^2\times\mathrm{K3})/\mathbb{Z}_2$, where $\mathbb{Z}_2$
is a freely acting involution. Its holonomy group equals
$\mathrm{SU}(2)\times\mathbb{Z}_2$, which implies that the type-II
string compactification is described by an effective four-dimensional
theory with $N=2$ supersymmetry. The Enriques Calabi-Yau is
self-mirror with Hodge numbers $h^{(2,1)}= h^{(1,1)}=11$, so that its
Euler number $\chi$ vanishes, and the massless sector of the
four-dimensional theory comprises 11 vector supermultiplets, 12
hypermultiplets and the $N=2$ graviton supermultiplet. In what follows
we concentrate on the vector multiplet sector, whose classical moduli
space, which is not affected by quantum corrections, equals the
special-K\"ahler space,
\begin{equation}
  \label{eq:4vector-special-K}
  \mathcal{M}_{\mathrm{vector}}=
  \frac{\mathrm{SL}(2)}{\mathrm{SO}(2)}\times
  \frac{\mathrm{O}(10,2)}{\mathrm{O}(10)\times\mathrm{O}(2)}\,.
\end{equation}
Its two factors are associated with $\mathrm{T}^2/\mathbb{Z}_2$ and the
$\mathrm{K3}$ fiber, and the special coordinates for these two spaces
will be denoted by $S$ and $T^a$, respectively.\footnote{%%%%%%%%%%%%
  The hypermultiplet moduli space contains the type-II dilaton and is
  of no concern to us. Its classical moduli space is given by the
  quaternion-K\"ahler space
  $\mathrm{O}(12,4)/[\mathrm{O}(12)\times\mathrm{O}(4)]$, as follows
  from the c-map \cite{Ferrara:1995yx}. } %%%%%%%%%%%%%%%%%%%%%%%%%%
In the limit $(S+\bar S)\to \infty$ one recovers the perturbative
result of the dual realization on the corresponding heterotic string
orbifold.

Obviously the classical moduli space \eqref{eq:4vector-special-K} is
invariant under the continuous group
$\mathrm{SL}(2)\times\mathrm{O}(10,2)$. However, at the quantum level
the model is invariant under the product of two discrete groups, namely the
$\Gamma(2)$ subgroup of $\mathrm{SL}(2;\mathbb{Z})$, and the group 
$\mathrm{O}(10,2;\mathbb{Z})$.  These groups must be realized as the
invariance group of a more complete effective field theory
description. We will call those the S- and T-duality groups,
respectively, although this nomenclature is not quite appropriate in
the type-II context.

Another model with 8 supersymmetries is the STU model
\cite{Sen:1995ff,Gregori:1999ns}, which may be regarded as a
truncation of the FHSV model, based on 3 vector multiplets and 4
hypermultiplets. Note that the STU model is also self-mirror and has
$\chi=0$. Its corresponding special-K\"ahler space equals,
\begin{equation}
  \label{eq:4vector-special-K-STU}
  \mathcal{M}_{\mathrm{vector}}=
  \frac{\mathrm{SL}(2)}{\mathrm{SO}(2)}\times
  \frac{\mathrm{SL}(2)}{\mathrm{SO}(2)}\times
  \frac{\mathrm{SL}(2)}{\mathrm{SO}(2)}\;.
\end{equation}
The duality group of this model is the product of the discrete $\Gamma(2)$ 
subgroups of each of the three $\mathrm{SL}(2)$ groups. 

For reasons of comparison we will also consider the so-called CHL
models \cite{Chaudhuri:1995fk}, which are invariant under 16
supersymmetries and whose S-dualities belong to the $\Gamma_1(\tilde
N)$ subgroup of $\mathrm{SL}(2;\mathbb{Z})$.  Here $\tilde N$ is an
integer parameter and the models with $\tilde N= 1, 2,3, 5, 7$, have
been studied in the literature \cite{Jatkar:2005bh}. The case $\tilde
N=1$ corresponds to the toroidal compactification of heterotic string
theory. The rank of the gauge group (corresponding to the number of
abelian gauge fields in the effective supergravity action) is then
equal to $r= 28, 20, 16, 12$ or $10$, respectively, and the
corresponding number of $N=2$ matter vector supermultiplets equals
$n=48/(\tilde N+1)-1$. Many of the studies of BPS black holes in CHL
models have been carried out based on an effective $N=2$ supergravity
description.

Let us now consider the underlying holomorphic function
$F(Y,\Upsilon)$ in terms of which the Wilsonian action for the vector
multiplet sector is encoded. As explained in the previous section the
dependence on the field $\Upsilon$ induces the presence of certain
higher-order derivative interactions, which, among others, involve the
square of the Weyl tensor. The definition of the $n+1$ complex fields
$Y^I$ was also discussed in the previous section.\footnote{%%%%%%%%%
  See footnote \ref{footnote-XY}. Note that $\Upsilon$ has been
  subject to a similar rescaling. }  %%%%%%%%%%%%%%%%%%%%%%%%%%%%%%%
The number $n$ will depend on the particular model that one is
considering. For example, the FHSV model and the STU model have $n=11$
and $n=3$, respectively.  Usually one assumes that the function can be
expanded in positive powers of $\Upsilon$. For type-II
compactifications on Calabi-Yau three-folds that are $\mathrm{K3}$
fibrations, the expansion takes the form
\begin{equation}
  \label{eq:F-expansion}
  F(Y,\Upsilon) = - \frac{Y^1 Y^a\eta_{ab}Y^b}{Y^0} +
  \sum_{g=1}^\infty\, \Upsilon^g\, F^{(g)}(Y)  \,, 
\end{equation}
where $a,b=2,\ldots,n$, and the symmetric matrix $\eta_{ab}$ is an
$\mathrm{SO}(n-2,1)$ invariant metric of indefinite signature.
Obviously this expression can be parametrized by
\begin{equation}
  \label{eq:top-string}
  F(Y,\Upsilon) =  \mathrm{i} (Y^0)^2 \, S\,T^a\eta_{ab}T^b +
  \Upsilon\, F^{(1)}(S,T) + \sum_{g=2}^\infty\,
  \frac{\Upsilon^g}{(Y^0)^{2g-2}} \, F^{(g)}(S,T)  \,, 
\end{equation}
where 
\begin{equation}
  \label{eq:special-coordinates}
  S=-\mathrm{i}\,Y^1/Y^0\,,\qquad T^a=-\mathrm{i}\,Y^a/Y^0\,,
\end{equation}
denote the special coordinates that parametrize the moduli space of
the Calabi-Yau three-folds. We stress that the classical moduli space
described by the first term of~\eqref{eq:top-string} is exact for the
models that we discuss in this paper. The function $F(Y,\Upsilon)$
takes the form of a loop expansion with $Y^0$ as a loop-counting
parameter. This is the form that is used for the topological string
where $Y^0$ is regarded as the inverse topological string coupling
constant and the functions $F^{(g)}(S,T)$ are the genus-$g$ twisted
partition functions.\footnote{ %%%%%%%%%%%%%%%%%%%%%%%%%%%%%%
  Hence $ F^{(g)}(Y) = (Y^0)^{-2g+2} \, F^{(g)}(S,T)$; when refering
  to the genus-$g$ partition functions in the text, we usually do not
  make a distinction between $F^{(g)}(Y)$ and $F^{(g)}(S,T)$. 
} %%%%%%%%%%%%%%%%%%%%%%%%%%%%%%%%%%%%%%%%%%%%%%%%%%%%%%%%%%%
The latter acquire non-holomorphic corrections
encoded by the holomorphic anomaly equation, whose structure is such
that the holomorphic dependence on the topological string coupling
constant is preserved \cite{Bershadsky:1993cx}.

On the other hand it is well known that non-holomorphic corrections
are also required to realize the relevant symmetries of the effective
action \cite{Dixon:1990pc}. In this context the holomorphic
contributions encode the Wilsonian effective action which is supposed
to arise from integrating out massive degrees of freedom.  The
Wilsonian action does not necessarily reflect all the symmetries of
the theory and those are recovered upon including the contributions
from the massless fields. These contributions contain non-holomorphic
terms. As we mentioned already in section \ref{sec:introduction}, it
turns out that the non-holomorphic corrections to the effective action
are not quite identical to the non-holomorphic contributions to the
genus-$g$ partition functions of the topological string, at least for
$g>1$. This will be further discussed in section
\ref{sec:non-holom-corr}.

Equivalence classes of the holomorphic function $F(Y,\Upsilon)$ are
governed by $\mathrm{Sp}(2n+2,\mathbb{Z})$ rotations of the
$2(n+1)$-component period vector of the underlying Calabi-Yau
holomorphic three-form, corresponding to $(Y^I,F_I)$, where
$F_I=(F_0,F_1,F_a)$ denotes the derivatives of $F$ with respect to
$Y^0$, $Y^1$ and $Y^a$, respectively. For the models based on
\eqref{eq:top-string} the invariance group is embedded into an
$\mathrm{SL}(2;\mathbb{Z})\times\mathrm{O}(n-1,2;\mathbb{Z})$ subgroup
of these monodromy transformations. In this section it is not
necessary to precisely specify this embedding. At the classical level,
where one retains only the first term in \eqref{eq:top-string}, the
continuous version of these monodromy transformations generate the
isometries of the moduli spaces. At the level of the four-dimensional
effective action these transformations are accompanied by
electric/magnetic duality transformations.

The period vector $(Y^I,F_I)$ plays a central role in the so-called
attractor equations for BPS black holes, which express their imaginary
parts (taken at the black hole horizon) in terms of the black hole
charges (c.f. \eqref{eq:attractor}).  Rather than concentrating on the
properties of the function \eqref{eq:top-string}, we will therefore
focus attention on the properties of this period vector. On the period
vector, invariance transformations are characterized by the fact that
the variations of the $Y^I$ induce the action on the $F_I(Y,\Upsilon)$
according to the monodromy matrix that also acts on the black hole
charges. Because the BPS attractor equations require $\Upsilon$ to
take a specific value at the horizon (namely $\Upsilon=-64$), it is
possible that the invariance arguments are not valid for arbitrary
$\Upsilon$. Based on previous work, it seems at least necessary to
restrict $\Upsilon$ to a real number. However, in this section this
aspect does not yet play a role. Furthermore, because the action of
the monodromies on the charges is not subject to corrections, the
action of the symmetry on the period vector must remain unchanged upon
introducing non-holomorphic corrections.

In view of the above it is of interest to define the monodromies
associated with the group
$\mathrm{SL}(2;\mathbb{Z})\times\mathrm{O}(n-1,2;\mathbb{Z})$, a
subgroup of which is expected to leave the model invariant. The action
of the S-duality group is defined as follows,
\begin{equation}
    \label{eq:S-duality}
    \begin{array}{rcl}
      Y^0 &\to& d \, Y^0 + c\, Y^1 \;,\\
      Y^1 &\to& a \, Y^1 + b \, Y^0 \;,\\
      Y^a &\to& d\, Y^a - \ft12 c \,\eta^{ab}\,F_b  \;,
    \end{array}
    \quad
    \begin{array}{rcl}
      F_0 &\to& a\,  F_0 -b\,F_1 \;, \\
      F_1 &\to& d \,F_1 -c\, F_0 \;,\\
      F_a &\to& a \, F_a -2b\, \eta_{ab}\, Y^b \;,
    \end{array}
\end{equation}
where $a,b,c,d$ are integer-valued parameters that satisfy $ad-bc=1$
which parametrize (a subgroup of) $\mathrm{SL}(2;\mathbb{Z})$.

For the T-duality group, general transformations are most easily
generated by products of a number of specific finite transformations.
Those transformations that belong to the
$\mathrm{O}(n-2,1;\mathbb{Z})$ subgroup are manifest in the above
description and do not have to be considered. Then there are $n-1$
abelian transformations generated by
\begin{equation}
  \label{eq:shifs}
    \begin{array}{rcl}
      Y^0 &\to& Y^0 \,,\\
      Y^1 &\to& Y^1 \,,\\
      Y^a &\to& Y^a -\lambda^a \,Y^0\,,
    \end{array} 
    \quad
    \begin{array}{rcl}
      F_0 &\to& F_0 +\lambda^a F_a + \lambda^a\eta_{ab}\lambda^b \,Y^1
      \,,\\ 
      F_1 &\to& F_1 +2\,\lambda^a \eta_{ab} Y^b -
      \lambda^a\eta_{ab}\lambda^b \,Y^0 \,,\\
      F_a &\to& F_a +2\,\eta_{ab}\lambda^b \,Y^1 \,,
  \end{array}
\end{equation}
where the $\lambda^a$ are integers. Finally the full
$\mathrm{O}(n-1,2;\mathbb{Z})$ group is generated provided one also
includes the following transformation,
\begin{equation}
    \label{eq:T-inversion}
    \begin{array}{rcl}
      Y^0 &\to& F_1 \;,\\
      Y^1 &\to& - F_0  \;,\\
      Y^a &\to&  Y^a  \;,
    \end{array}
    \quad
    \begin{array}{rcl}
      F_0 &\to& -Y^1 \;, \\
      F_1 &\to& Y^0 \;, \\
      F_a &\to& F_a \;.
    \end{array}
\end{equation}
Observe that the square of this transformation equals the identity.

In the case that the higher-order genus terms in \eqref{eq:top-string}
are suppressed, it is straightforward to evaluate the behaviour of
these transformations on the special coordinates $S$ and $T^a$. Under
S-duality we find the well-known results,
\begin{equation}
  \label{eq:ST-Sdual}
  S \rightarrow
  \frac{a\,S-\mathrm{i}b}{\mathrm{i}c\,S+d} \;, 
  \qquad T^a\to T^a\;.
\end{equation}
The T-duality transformations \eqref{eq:shifs} and
\eqref{eq:T-inversion} lead to, respectively, 
\begin{equation}
  \label{eq:T-S-Sdual}
  S\to S\;, \qquad T^a\to T^a+\mathrm{i}
  \,\lambda^a \;, \qquad T^a\to 
  \frac{T^a}{T^b\eta_{bc}T^c}\;. 
\end{equation}
However, these S- and T-duality transformations become much more
complicated in the presence of higher-genus contributions in
\eqref{eq:top-string}. Insisting on the same symmetry (i.e.,
characterized by the same monodromy matrix) will restrict these
higher-genus contributions. This was demonstrated, for instance in
\cite{LopesCardoso:1999ur}, in a simpler situation.

In what follows we concentrate on the periods and thus consider
holomorphic derivatives of the function $F$, which is itself not
holomorphic, 
\begin{equation}
  \label{eq:het-F}
  F = - \frac{Y^1\,Y^a\eta_{ab}Y^b}{Y^0} + 2\mathrm{i} 
  \,\Omega(Y,\bar Y,\Upsilon,\bar\Upsilon)\;,
\end{equation}
where $\Omega$ encodes the non-classical contributions in accordance
with \eqref{eq:F-decomposition}. We will still be assuming that
$\Omega$ depends only on positive powers of $\Upsilon$ and
$\bar\Upsilon$ compensated by negative even powers of $Y^0$ and/or
$\bar Y^0$ so as to make \eqref{eq:het-F} homogeneous of second degree
(but not necessarily holomorphic). Furthermore we expect that $\Omega$
vanishes for $\Upsilon=0$. In the case studied before
\cite{Cardoso:2004xf}, where the $F^{(g)}$ vanish for $g>1$, it turns
out that $\Omega$ could be written as a real function.  As long as
$\Omega$ is harmonic, which implies that it can be written as the
difference of a holomorphic and an anti-holomorphic function, this
modification has no consequences when considering the periods, as the
latter will remain holomorphic. Irrespective of these precise
properties the $F_I$ can be written as follows,
\begin{eqnarray}
  \label{eq:F_I}
  F_0&=& \frac{Y^1}{(Y^0)^2}\,Y^a\eta_{ab}Y^b -\frac{2\mathrm{i}}{Y^0}
  \left[-Y^0 \frac{\partial}{\partial Y^0}+
  S\frac{\partial}{\partial S}+ T^a\frac{\partial}{\partial T^a}
  \right]\Omega \;,  \nonumber\\ 
  F_1&=&{}- \frac{1}{Y^0}\,Y^a\eta_{ab}Y^b +\frac{2}{Y^0} \,
  \frac{\partial\Omega}{\partial S} \;,\nonumber\\
  F_a&=& {}- 2 \frac{Y^1}{Y^0}\,\eta_{ab}Y^b +\frac{2}{Y^0} \,
  \frac{\partial\Omega}{\partial T^a} \;,
\end{eqnarray}
where we regard $\Omega$ as a function of $Y^0$, $S$ and $T^a$ (and
possibly their complex conjugates). 

With these results the S-duality transformations \eqref{eq:S-duality}
take the form, 
\begin{eqnarray}
  \label{eq:full-S}
  Y^0&\to& \Delta_{\mathrm{S}}\, Y^0\;, \nonumber\\
  Y^1&\to& a\,Y^1+ b\,Y^0\;, \nonumber\\
  Y^a&\to& \Delta_{\mathrm{S}}\, Y^a - \frac{c}{Y^0} \,\eta^{ab}\,
  \frac{\partial\Omega}{\partial T^b}  \;,  
\end{eqnarray}
with 
\begin{equation}
  \label{eq:Delta-S}
  \Delta_{\mathrm{S}} = d+\mathrm{i}c\,S\,.
\end{equation}
On the special coordinates $S$ and $T^a$ these transformations extend
the previous result \eqref{eq:ST-Sdual},
\begin{equation}
  \label{eq:ST-S-full}
  S \rightarrow
  \frac{a\,S-\mathrm{i}b}{\mathrm{i}c\,S+d} \;, 
  \qquad T^a\to T^a +\frac{\mathrm{i} c}{\Delta_{\mathrm{S}}\,(Y^0)^2}
  \,\eta^{ab} 
  \,\frac{\partial\Omega}{\partial T^b}  \;,
\end{equation}
and we note the useful relations
\begin{equation}
  \label{eq:dS-dS}
  \frac{\partial{S^\prime}}{\partial{S}}
  =\Delta_{\mathrm{S}}{}^{-2}\,,\qquad \frac1{S+\bar S}\to
  \frac{\vert\Delta_{\mathrm{S}}\vert^2}{S+\bar S} = 
  \frac{\Delta_{\mathrm{S}}{}^2}{S+\bar S}
  -\mathrm{i} c\,\Delta_{\mathrm{S}}\,.  
\end{equation}

Assuming that the above transformations constitute an invariance of
the model, we require that the S-duality transformations of the $Y^I$
induce the expected transformations of the $F_I$ upon substitution.
This leads to the following result,\footnote{
$(O)^\prime_{\mathrm{S,T}}$
denotes the change of $O$ under S- or T-duality induced by the 
transformation of all the arguments on which $O$ depends.}
\begin{eqnarray}
  \label{eq:S-invariance}
  \left(\frac{\partial\Omega}{\partial T^a}\right)^\prime_\mathrm{S} &=&
  \frac{\partial\Omega}{\partial T^a} \;, \nonumber\\
  \left(\frac{\partial\Omega}{\partial S}\right)^\prime_\mathrm{S} -
  \Delta_{\mathrm{S}}{}^2\,\frac{\partial\Omega}{\partial S} &=&
  \frac{\partial(\Delta_{\mathrm{S}}{}^2)}{\partial{S}} 
  \left[-\tfrac12 Y^0 \frac{\partial\Omega}{\partial Y^0}
  -\frac{\mathrm{i}c}{4\,\Delta_{\mathrm{S}}\,(Y^0)^2}
  \,\frac{\partial\Omega}{\partial T^a}\eta^{ab}
  \frac{\partial\Omega}{\partial T^b} \right] \;,
  \nonumber\\ 
  \left(Y^0\frac{\partial\Omega}{\partial Y^0}\right)^\prime_\mathrm{S} &=&
  Y^0 \frac{\partial\Omega}{\partial Y^0} 
  +\frac{\mathrm{i}c}{\Delta_{\mathrm{S}}\,(Y^0)^2}
  \,\frac{\partial\Omega}{\partial T^a}\eta^{ab}
  \frac{\partial\Omega}{\partial T^b}\;. 
\end{eqnarray}
It is instructive to consider the consequences of these equations in
case that the dependence on the $T$-moduli is suppressed (i.e., 
$\partial\Omega/\partial T^a=0$) and non-holomorphic terms are absent
(so that we may use the decomposition \eqref{eq:top-string}). The
result is that the functions $F^{(g)}(S,T)$ are modular forms of weight
$2g-2$, as the above equations take the form, 
\begin{eqnarray}
  \label{eq:holo-S}
  \partial_SF^{(1)}(S,T)&\longrightarrow& \Delta_{\mathrm{S}}^{2}\,
  \partial_S F^{(1)}(S,T)  \,, \nonumber \\ 
  F^{(g)}(S,T)&\longrightarrow& \Delta_{\mathrm{S}}^{2g-2}
  F^{(g)}(S,T)\,,\quad(g>1) \nonumber\\ 
  D_SF^{(g)}(S,T)   &\longrightarrow&  \Delta_{\mathrm{S}}^{2g} D_S
  F^{(g)}(S,T)\,,   \quad(g>1) 
\end{eqnarray}
where $D_SF^{(g)}(S,T) \equiv [\partial_S-2(g-1)
\partial_S\ln\eta^2]F^{(g)}(S,T)$ with $\eta(S)$ the Dedekind function.
Here $\partial_S\ln\eta^2$ acts as a connection, in view of its
transformation law,
\begin{equation}
  \label{eq:s-connection}
      \partial_S\ln \eta^2 \to\Delta_{\mathrm{S}}^2 \,\partial_S\ln
  \eta^2 +\tfrac12 \partial_S\Delta_{\mathrm{S}}^2\,,
\end{equation}
but alternative connections exist that will lead to identical results.
In the holomorphic case the first derivative with respect to
$\Upsilon$ of $\Omega$ is known to be an invariant function
\cite{deWit:1996ix}, and this is consistent with the second equation
of \eqref{eq:holo-S}.

The same reasoning applies to T-duality. Under the transformation
\eqref{eq:shifs} it follows from \eqref{eq:F_I} that all the
derivatives $\partial\Omega/\partial Y^0$, $\partial\Omega/\partial S$
and $\partial\Omega/\partial T^a$ must be invariant under integer
shifts $T^a\to T^a + \mathrm{i}\,\lambda^a$. 
For the T-duality transformation \eqref{eq:T-inversion} the analysis
is more subtle. Using \eqref{eq:F_I} we derive,
\begin{eqnarray}
  \label{eq:full-T}
  Y^0&\to& \Delta_{\mathrm{T}}\, Y^0\;, \nonumber\\
  Y^1&\to& \Delta_{\mathrm{T}}\, Y^1 + \frac{2\mathrm{i}}{Y^0}
  \left[-Y^0\frac{\partial\Omega}{\partial Y^0} +
  T^a\frac{\partial\Omega}{\partial T^a}\right]\;, \nonumber\\
  Y^a&\to& Y^a \;,
\end{eqnarray}
with 
\begin{equation}
  \label{eq:Delta-T}
  \Delta_{\mathrm{T}} = T^a\eta_{ab}T^b +\frac{2}{(Y^0)^2}
  \frac{\partial\Omega}{\partial S}\;. 
\end{equation}
On the special coordinates the transformation \eqref{eq:full-T}
extends the previous result \eqref{eq:T-S-Sdual},
\begin{eqnarray}
  \label{eq:ST-T-full}
  S&\rightarrow&S+ \frac2{\Delta_{\mathrm{T}}(Y^0)^2}
  \,\left[- Y^0 \frac{\partial\Omega}{\partial Y^0} + 
  T^a\frac{\partial\Omega}{\partial T^a} \right]  \;,
  \nonumber\\ 
  \qquad T^a&\to& \frac{T^a}{\Delta_{\mathrm{T}}}  \;.
\end{eqnarray} 
Again we assume that the above transformations constitute an
invariance of the model, and require that the T-duality transformation
\eqref{eq:full-T} of the $Y^I$ induces the expected transformations of
the $F_I$ upon substitution. This leads to
\begin{eqnarray}
  \label{eq:T-invariance}
  \left(\frac{\partial\Omega}{\partial S}\right)^\prime_\mathrm{T} &=& 
  \frac{\partial\Omega}{\partial S} \;, \nonumber\\
  \left(\frac{\partial\Omega}{\partial T^a}\right)^\prime_\mathrm{T} &=&
  \left(\Delta_{\mathrm{T}} \,\delta_a{}^b - 2\,\eta_{ac}T^c
  T^b\right) \,\frac{\partial\Omega}{\partial T^b}
  +2\,\eta_{ab}T^b\;Y^0 \frac{\partial\Omega}{\partial Y^0} \;,
  \nonumber\\ 
  \left(Y^0 \frac{\partial\Omega}{\partial Y^0}\right)^\prime_\mathrm{T} &=&
  Y^0 \frac{\partial\Omega}{\partial Y^0} + 
  \frac{4}{\Delta_{\mathrm{T}}\,(Y^0)^2}
  \,\frac{\partial\Omega}{\partial S} 
  \left[- Y^0 \frac{\partial\Omega}{\partial Y^0} +
  T^a\frac{\partial\Omega}{\partial T^a}\right]\;. 
\end{eqnarray}
To appreciate the first term on the right-hand side of the second
equation we note 
\begin{equation}
  \label{eq:dT-dT}
  \frac{\partial{T^{\prime a}}}{\partial{T^b}}=
  \frac1{\Delta_{\mathrm{T}}} \left[\delta^a{}_b -
  \frac{2\,T^a \,\eta_{bc}T^c}{\Delta_{\mathrm{T}}} -
  \frac{2\,T^a}{\Delta_{\mathrm{T}}(Y^0)^2}\,
  \frac{\partial^2\Omega}{\partial{T^b}\partial{S}} 
  \right] \;.
\end{equation}
In case that the $S$-dependence is suppressed so that we can drop the
terms proportional to $\partial_S\Omega$, \eqref{eq:dT-dT} is
precisely the inverse of the term appearing in the second equation
\eqref{eq:T-invariance}.

As before it is instructive to consider the consequences of these
equations in case that non-holomorphic terms are absent (so that we
use the decomposition \eqref{eq:top-string}), assuming this time that
the dependence on the $S$ modulus can be ignored, so that
$\partial_S\Omega=0$. The result is that the $F^{(g)}(S,T)$ are
holomorphic automorphic forms of weight $2g-2$, as the above
equations reduce to (note that $\Delta_{\mathrm{T}}= T^a\eta_{ab}T^b$
in this case),
\begin{eqnarray}
  \label{eq:holo-T}
  \partial_{T^a} F^{(1)}(S,T) &\longrightarrow&
  \left(\Delta_{\mathrm{T}} \,\delta_a{}^b - 2\,\eta_{ac}T^c 
  T^b\right) \,\partial_{T^b} F^{(1)}(S,T)  \,, \nonumber \\ 
  F^{(g)}(S,T)&\longrightarrow& \Delta_{\mathrm{T}}^{\;2g-2}
  F^{(g)}(S,T)\,,\qquad(g>1) \nonumber\\ 
  D_{T^a}F^{(g)}(S,T) &\longrightarrow& 
  \left(\Delta_{\mathrm{T}} \,\delta_a{}^b - 2\,\eta_{ac}T^c 
  T^b\right) \,
 \Delta_{\mathrm{T}}^{\;2g-2} D_{T^b} F^{(g)}(S,T)\,,
  \quad(g>1) 
\end{eqnarray}
where $D_{T^a}F^{(g)}(S,T) \equiv
[\partial_{T^a}+(g-1)\partial_{T^a}\ln\Delta_{\mathrm{T}}]
F^{(g)}(S,T)$. Again this result is consistent with the fact that the first
derivative with respect to $\Upsilon$ must be an invariant function in
the holomorphic case. Here we made use of a connection
$-\tfrac12\partial_T\ln\Delta_{\mathrm{T}}$, as
\begin{equation}
  \label{eq:t-connection}
  -\tfrac12\partial_{T^a}\ln\Delta_{\mathrm{T}} \to
  \left(\Delta_{\mathrm{T}} \,\delta_a{}^b - 2\,\eta_{ac}T^c
  T^b\right) \,\left[-\tfrac12\partial_{T^b}\ln\Delta_{\mathrm{T}} +
  \partial_{T^b}\ln\Delta_{\mathrm{T}} \right]\,. 
\end{equation}
However, other (less trivial) connections are possible. For instance,
in the FHSV model one may use $\tfrac14\partial_T\ln\Phi(T)$, where
$\Phi(T)$ is the holomorphic automorphic form of weight 4
(c.f.\eqref{eq:Phi}). A non-holomorphic connection is given by
$-\partial_{T}\ln[(T+\bar T)^a\eta_{ab}(T+\bar T)^b]$, which is
invariant under imaginary shifts of the $T^a$. Note that, in the same
approximation as above, the T-duality transformation \eqref{eq:full-T}
acts as
\begin{equation}
  \label{eq:re-T-square}
  (T+\bar T)^a\eta_{ab}(T+\bar T)^b \to
  \frac{1}{\vert\Delta_\mathrm{T}\vert^{2}} \, 
  (T+\bar T)^a\eta_{ab}(T+\bar T)^b\,. 
\end{equation}
We refer to \cite{Grimm:2007tm} for further discussion.

Returning to the more general case it follows that both
$\partial_S\Omega$ and $Y^0\frac{\partial\Omega}{\partial Y^0}-2\,S
\frac{\partial\Omega}{\partial S}$ are T-duality invariant, whereas
$\partial_{T^a}\Omega$ is S-duality invariant. Furthermore, the
combination $Y^0 \frac{\partial\Omega}{\partial Y^0}-
T^a\frac{\partial\Omega}{\partial T^a}$ turns out to be invariant
under S-duality, while, under the T-duality \eqref{eq:full-T}, it is
invariant up to a sign change. We also note the relations,
\begin{eqnarray}
    \label{eq:Delta-TS}
    \Delta_{\mathrm{T}}&\stackrel{\mathrm{T}}{\longrightarrow}&
    \frac1{\Delta_{\mathrm{T}}}\,,  \nonumber\\
    \Delta_{\mathrm{T}}&\stackrel{\mathrm{S}}{\longrightarrow}&
    \Delta_{\mathrm{T}} + 
    \frac{2\,\mathrm{i}\,c}{\Delta_{\mathrm{S}}\,(Y^0)^2}
    \left[-Y^0 \frac{\partial\Omega}{\partial Y^0}
      + T^a\frac{\partial\Omega}{\partial
    T^a}\right]  \,,  
    \nonumber\\
    \Delta_{\mathrm{S}}&\stackrel{\mathrm{T}}{\longrightarrow}&
    \Delta_{\mathrm{S}} + 
    \frac{2\,\mathrm{i}\,c}{\Delta_{\mathrm{T}}\,(Y^0)^2}  
    \left[- Y^0 \frac{\partial\Omega}{\partial Y^0} +
    T^a\frac{\partial\Omega}{\partial T^a}\right]  \,. 
\end{eqnarray}

This completes the review of S- and T-duality transformations in the
FHSV model and in similar models, such as the STU model.  We stress
once more that the central results, \eqref{eq:S-invariance} and
\eqref{eq:T-invariance}, hold in the presence of non-holomorphic
modifications. Furthermore, it should be clear that $\Omega$ is not an
invariant function. While the fields $\Upsilon$ and $\bar\Upsilon$ do
not enter explicitly into the monodromies \eqref{eq:S-duality},
\eqref{eq:shifs} and \eqref{eq:T-inversion}, the corresponding
transformations induced on $Y^0$, $S$, and $T^a$ depend in a
complicated way on $\Upsilon$ and $\bar\Upsilon$. In the next two
sections we will discuss how to solve these equations iteratively in
$\Upsilon=\bar\Upsilon$. In section \ref{sec:measure-factor-Omega-1},
we restrict ourselves to terms linear in $\Upsilon=\bar\Upsilon$ with
the aim of studying the subleading corrections to the mixed black hole
partition function. These terms coincide with the genus-1 partition
functions of the topological string.  Then, in subsection
\ref{sec:about-Omega-2}, we analyse higher-order terms in
$\Upsilon=\bar\Upsilon$, related to the genus-2 partition function of
the topological string. As we intend to demonstrate the result no
longer agrees directly with the topological string. The underlying
reason for this different result resides in the fact that the
transformation rules depend on $\Upsilon,\bar\Upsilon$, unlike in the
case of the topological string.

%%%%%%%%%%%%%%%%%%%%%%%%%%%%%%%%%%%%%%%%%%%%%%%%%%%%%%%%%%%%%%%%%%%
\section{The measure factor for the mixed partition function}
\label{sec:measure-factor-Omega-1}
\setcounter{equation}{0}
%%%%%%%%%%%%%%%%%%%%%%%%%%%%%%%%%%%%%%%%%%%%%%%%%%%%%%%%%%%%%%%%%%%

The consequences of the duality symmetry, which are expressed by the
equations \eqref{eq:S-invariance} and \eqref{eq:T-invariance} for the
function $\Omega$ defined in \eqref{eq:het-F}, can be studied by
iteration in powers of $\Upsilon$ and $\bar\Upsilon$. Therefore it is
convenient to expand $\Omega$ as follows,
\begin{equation}
  \label{eq:exp-Omega}
 \Omega(Y,\bar Y,\Upsilon,\bar\Upsilon) = \sum_{g=1}^\infty\;
 \Omega^{(g)}(Y,\bar Y,\Upsilon,\bar\Upsilon) \,,
\end{equation}
where $\Omega^{(g)}$ may in general contain various monomials in
$\Upsilon$ and $\bar \Upsilon$ of degree $g$. As $\Omega^{(g)}$ must
be a real function that is homogeneous of degree two, the coefficients
of these monomials take the form of functions of $S$ and $T^a$, as
well as of their complex conjugates, divided by homogeneous
polynomials of $Y^0$ and $\bar Y^0$ of degree $2(g-1)$. In particular
$\Omega^{(1)}(S,T)$ is known for a large variety of models. 

In the context of large black holes, only $\Omega^{(1)}(S,T)$ is
expected to contribute to the mixed partition function
\eqref{eq:partition1-osv} in the semi-classical approximation, as
discussed at the end of section \ref{sec:partition-entropy-function}.
Therefore we restrict ourselves here to the case $g=1$. This result
will enable us to evaluate the effective measure factor for the mixed
partition function at the end of this section.

We study the constraints imposed by S- and T-duality invariance for
the terms linear in $\Upsilon$ and/or $\bar\Upsilon$, and their
non-holomorphic corrections, proceeding by iteration and assuming that
the duality invariance will be realized order-by-order in $\Upsilon$
(subject to $\bar\Upsilon=\Upsilon$). We consider both the FHSV and
STU models, which have $N=2$ supersymmetry, as well as the $N=4$
supersymmetic CHL models. Considering this variety of models will be
helpful in callibrating the normalization of $\Omega$. All these
models share the property that the first term in \eqref{eq:top-string}
is not modified by quantum corrections. In this iterative procedure
the term $\Omega^{(1)}$, which is linear in $\Upsilon$ or
$\bar\Upsilon$, is subject to relatively simple equations,
\begin{eqnarray}
  \label{eq:F_1}
  \frac{\partial\Omega^{(1)}}{\partial T^a} 
  &\stackrel{\mathrm{S}}{\longrightarrow}&
  \frac{\partial\Omega^{(1)}}{\partial T^a} \;, \nonumber\\
  \frac{\partial\Omega^{(1)}}{\partial S}
  &\stackrel{\mathrm{S}}{\longrightarrow}& 
  \Delta_{\mathrm{S}}^{\;2}\,\frac{\partial\Omega^{(1)}}{\partial S}\;,  
  \nonumber\\ 
  \frac{\partial\Omega^{(1)}}{\partial S}
  &\stackrel{\mathrm{T}}{\longrightarrow}& 
  \frac{\partial\Omega^{(1)}}{\partial S} \;, \nonumber\\
  \frac{\partial\Omega^{(1)}}{\partial T^a}
  &\stackrel{\mathrm{T}}{\longrightarrow}& 
  \left(\eta_{cd}T^cT^d \,\delta_a{}^b - 2\,\eta_{ac}T^c
  T^b\right) \,\frac{\partial\Omega^{(1)}}{\partial T^b}\;. 
\end{eqnarray}
These equations are obviously satisfied by assuming that $\Omega^{(1)}$ is
the sum of an S-duality invariant function of $S$, and a T-duality
invariant function of $T^a$. Such invariant modular and automorphic
functions are usually quite rare, so that invariance under the duality
group will pose strong restrictions. 

The solutions of the above equations are known for the FHSV model,
where the contribution linear in $\Upsilon$ or $\bar\Upsilon$ takes
the following form \cite{Harvey:1996ts,Klemm:2005pd},
\begin{eqnarray}
  \label{eq:Omega-FHSV}
  \Omega_{\mathrm{FHSV}}^{(1)}(S,\bar S,T, \bar T, \Upsilon,\bar\Upsilon)
  &=&{}
  \frac{1}{256\,\pi} \Big[\tfrac12 \Upsilon
  \ln[\eta^{24}(2S)\,\Phi(T)] 
  +\tfrac12\bar\Upsilon \ln[\eta^{24}(2\bar S)\,\Phi(\bar T)]
  \nonumber\\
  &&{}\hspace{1cm}
  +(\Upsilon+\bar \Upsilon) 
  \ln [(S+\bar S)^3 (T+\bar T)^a\eta_{ab}(T+\bar T)^b]\Big]
  \,.
\end{eqnarray} 
For real values of $\Upsilon$, this result is indeed invariant under
S-duality.\footnote{%%%%%%%%%%%%%%%%%%%%%%%%%%%%%%%%%%%%%%%%%%%%%%
  Here and in the following we make use of the modular transformation
  rule and the asymptotic expansion of the Dedekind eta function,
  \begin{eqnarray}
  \label{eq:eta-function}
  \ln \eta^{24}(S) &\to& \ln \eta^{24}(S) +12\,\ln \Delta_{\mathrm{S}} \,,
  \nonumber\\ 
  \ln\eta(S) &\approx& -\tfrac1{12} \pi S - \mathrm{e}^{-2\pi S}
  +\mathcal{O}(\mathrm{e}^{-4\pi S})\,.
  \end{eqnarray}
} %%%%%%%%%%%%%%%%%%%%%%%%%%%%%%%%%%%%%%%%%%%%%%%%%%%%%%%%%%%%%%%%
The S-duality transformations of this model constitute the $\Gamma(2)$
subgroup of $\mathrm{SL}(2;\mathbb{Z})$, defined by $a,d=
1\mod 2$ and $b,c = 0\mod 2$ in \eqref{eq:ST-Sdual}. The result is also
T-duality invariant in view of the fact that $\Phi(T)$ is a
holomorphic automorphic form of weight 4 \cite{Borcherds:1996},
\begin{equation}
  \label{eq:Phi}
  \Phi(T)= \prod_{r>0} \left(\frac{1-\mathrm{e}^{-2\pi\,r\cdot T}}
   {1+\mathrm{e}^{-2\pi\,r\cdot T}}\right)^{2c_1(r^2)} \,,
\end{equation}
transforming under the T-duality transformation \eqref{eq:full-T}
(suppressing the $S$-dependence) as
\begin{equation}
  \label{eq:Phi-T}
  \Phi(T)\to \Delta_{\mathrm{T}}^{\;4} \,\Phi(T)\,. 
\end{equation}
Indeed, \eqref{eq:Omega-FHSV} can be written as the sum of two
invariant functions, one of $S$ and $\bar S$ and one of 
$T^a$ and $\bar T^a$, respectively, which for large real values of $S$
and $T^a$ satisfies,  
\begin{equation}
  \label{eq:Omega-FHSV-large-TS}
  \Omega^{(1)}_\mathrm{FHSV}\approx - \frac{\Upsilon\,S +
    \bar\Upsilon\, \bar S}{128} \,. 
\end{equation}
It contains non-holomorphic terms, which are crucial for the duality
invariance, 
%for real values of $\Upsilon$, 
equal to
\begin{eqnarray}
  \label{eq:Omega-FHSV-nonholo}
  \Omega_{\mathrm{FHSV}}^{(1) \,\mathrm{nonholo}} = 
  \frac{\Upsilon+\bar \Upsilon}{256\,\pi} 
  \ln [(S+\bar S)^3 (T+\bar T)^a\eta_{ab}(T+\bar T)^b] \,. 
\end{eqnarray}
Observe that the duality invariance of
$\Omega^{(1)}_\mathrm{FHSV}$ is only realized 
for real values of $\Upsilon$.
Therefore we do not know a priori whether to write $\Upsilon$ or its
complex conjugate. The way in which this potential ambiguity has been
resolved, is by assuming that purely holomorphic terms are always
accompanied by a power of $\Upsilon$ and purely anti-holomorphic terms
by a power of $\bar\Upsilon$, whereas for the mixed terms we assign
$\Upsilon$ and $\bar\Upsilon$ such as to preserve the reality
properties of $\Omega$ for complex $\Upsilon$. 
At this point, it is not quite clear how this procedure
will work out at higher orders in $\Upsilon$ and $\bar\Upsilon$, but
we know from the explicit evaluation of $\Omega^{(2)}$ for the FHSV
model, which we will present in the next section, that no problems are
encountered.

Subsequently we turn to the STU model, based on the function
\begin{equation}
  \label{eq:F0-STU}
  F^{(0)}(Y)= -\frac{Y^1 Y^2 Y^3}{Y^0} =  \mathrm{i}\,(Y^0)^2 \, S T U\,,
\end{equation}
corresponding to $\eta_{12}=\eta_{21}=\ft12$ and
$\eta_{11}=\eta_{22}=0$. In this case, we have \cite{Gregori:1999ns}, 
\begin{eqnarray}
  \label{eq:Omega-STU}
  &&
  \Omega_{\mathrm{STU}}^{(1)}(S,\bar S,T,\bar T,U,\bar
  U,\Upsilon,\bar\Upsilon) =\nonumber\\
  &&\hspace{1cm}
  \frac{1}{256\,\pi} \Big[4\, \Upsilon
  \ln[\vartheta_2(S)\,\vartheta_2(T)\,\vartheta_2(U)] 
  + 4\,\bar\Upsilon
  \ln[\vartheta_2(\bar{S})\,\vartheta_2(\bar{T})\,\vartheta_2(\bar{U})] 
  \nonumber\\
  &&{}\hspace{2cm}
  +(\Upsilon+\bar \Upsilon) 
  \ln [(S+\bar S) (T+\bar T)(U+\bar U)]\Big]  \,,
\end{eqnarray} 
where 
\begin{equation}
  \label{eq:theta_2}
  \vartheta_2(S) = \frac{2\,\eta^2(2S)}{\eta(S)} \,. 
\end{equation}
For large real values of $S$, $T$ and $U$, this result yields
\begin{equation}
  \label{eq:Omega-STU-large-STU}
  \Omega^{(1)}_\mathrm{STU}\approx -
    \frac{\Upsilon(S+T+U)+\bar\Upsilon(\bar S+\bar T +\bar U)}{256} \,, 
\end{equation}
and its non-holomorphic contribution equals, 
\begin{equation}
  \label{eq:nonholo-STU}
  \Omega^{(1)\,\mathrm{nonholo}}_\mathrm{STU}= 
  \frac{\Upsilon+\bar\Upsilon}{256\,\pi} 
  \ln[(S+\bar{S})(T+\bar{T})(U+\bar{U})] \,.
\end{equation}
Assuming that the real part of $S$ is much larger than that of $T$ and
$U$, the two results \eqref{eq:Omega-FHSV-large-TS} and
\eqref{eq:Omega-STU-large-STU} coincide up to a factor 2. This is
related to the fact that the STU model has been defined on the type-II
side. The relation between the field $S$ and the heterotic dilaton
must involve a factor 2. When this is taken into account the two
results are in fact equal, in agreement with \cite{Aspinwall:1995vk}. 

It is instructive to confront some of the previous results with the
solution of the holomorphic anomaly equation for $\Omega^{(1)}$ for
generic Calabi-Yau compactifications,
\begin{eqnarray}
  \label{eq:F1-nonholo}
  4\pi\,\Omega^{(1)\,\mathrm{nonholo}}\Big\vert_{\Upsilon=-64} &=& -
  \tfrac12 \ln \left\vert\det 
  [\,\mathrm{Im} \,2\, F^{(0)}_{KL}]\right\vert  + \Big(\tfrac1{24}
  \chi -1 \Big) \ln \frac{K^{(0)}}{\vert Y^0\vert^2} \,, 
\end{eqnarray}
where we adjusted the proportionality constant to have agreement with
previous results. The quantity $K$ is generally defined by
\begin{equation}
  \label{eq:K-def}
  K = \mathrm{i}({\bar Y}^I F_I-Y^I {\bar F}_I ) \,.
\end{equation}
Here $F_I$ and $F_{IJ}$ refer to the derivatives of the general
function $F$ and may thus contain non-holomorphic contributions.
However, $F^{(0)}_{IJ}$ and $K^{(0)}$ refer only to the corresponding
expressions for $\Upsilon=0$, so that non-holomorphic terms are
absent. Then the K\"ahler potential $\mathcal{K}$ and the
determinant of the special-K\"ahler metric in the standard
representation \cite{de Wit:1984pk} (see also \cite{Ceresole:1995ca}),
are given by 
\begin{equation}
  \label{eq:K-pot-Omega-general}
  \mathcal{K}= - \ln[K^{(0)}/\vert Y^0\vert^2]\,,\qquad 
    g= -\,
    \mathrm{e}^{(n+1)\mathcal{K}}\,\det[\,\mathrm{Im}\,2\,F^{(0)}_{KL}]\,. 
\end{equation}

In the case at hand, where the function $F^{(0)}$ coincides with
\eqref{eq:het-F} in the $\Upsilon=0$ limit, the expression for the
K\"ahler potential and $\Omega^{(1)}$ are given by
\begin{eqnarray}
  \label{eq:KP-Omega-1}
  \mathcal{K}= -
  \ln[K^{(0)}/\vert Y^0\vert^2]&=&{} -\ln[(S+\bar S) (T+\bar
  T)^a\eta_{ab}(T+\bar T)^b] \,,\nonumber\\[.8ex]
  4\pi\,\Omega^{(1)\,\mathrm{nonholo}}\Big\vert_{\Upsilon=-64} &=& 
  \left(\frac{\chi}{24}-2-\frac{n-3}2\right)\ln(S+\bar{S})
  \nonumber\\ 
  &&{}
  +\left(\frac{\chi}{24}-2\right)\ln[(T+\bar{T})^a\eta_{ab}(T+\bar
  T)^b]  \,,
\end{eqnarray}
where we used the relation,
\begin{equation}
  \label{eq:det-F}
  \det[\mathrm{Im} \, 2\,F_{KL}^{(0)}]= 2^{n-1} (S+\bar
  S)^{n-3}\,\det[-\eta_{ab}]\;  
  \left[\frac{K^{(0)} }{\vert Y^0\vert^2}\right]^2  \,,
\end{equation}
which holds for the same class of functions.  For the Enriques
Calabi-Yau three-fold, $n=11$ and $\chi=0$, so that
\eqref{eq:F1-nonholo} coincides with \eqref{eq:Omega-FHSV-nonholo},
provided we set $\Upsilon=-64$. Similarly for the STU model, where one
has $\chi=0$ and $n=3$, the result coincides with
\eqref{eq:nonholo-STU}.

One may also consider the class of CHL models which have $N=4$
supersymmetry \cite{Chaudhuri:1995fk} and which we already mentioned
in section \ref{sec:omega-duality-constraints}.  These models are
invariant under the S-duality group $\Gamma_1(\tilde
N)\subset\mathrm{SL}(2;\mathbb{Z})$, which is generated by
\eqref{eq:ST-S-full} with the transformation parameters restricted to
$c = 0~\mod \tilde N$ and $a,d=1~\mod \tilde N$. They contain no
higher-genus contributions beyond genus-1. As discussed in
\cite{Jatkar:2005bh} the function $\Omega_k$ can be expressed in terms
of the unique cusp forms of weight $k+2$ associated with the S-duality
group $\Gamma_1(\tilde N)\subset\mathrm{SL}(2;\mathbb{Z})$, defined by
$f^{(k)}(S) = \eta^{k+2}(S)\,\eta^{k+2}(\tilde N S)$ where,
\begin{equation}
  \label{eq:f-S-dual}
  f^{(k)} (S^\prime) = \Delta_{\mathrm{S}}^{\,k+2}  \, f^{(k)}(S)\,.
\end{equation}
The result for $\Omega_k$ then takes the following form
\cite{LopesCardoso:2006bg},
\begin{equation}
  \label{eq:Omega-het}
  \Omega_k(S,\bar S,\Upsilon,\bar\Upsilon) =
  {}\frac{1}{256\,\pi} \Big[\Upsilon \ln f^{(k)}(S)  +
\bar\Upsilon \ln f^{(k)}(\bar S) + \ft12(\Upsilon+\bar \Upsilon)
\ln (S+\bar S)^{k+2} \Big] \,.
\end{equation}
Note that this result agrees with the terms obtained for the
corresponding effective actions (see, for instance,
\cite{Harvey:1996ir,Gregori:1997hi}).

For large real value of $S$ we obtain the same result
\eqref{eq:Omega-FHSV-large-TS} as for the FHSV model.  The
non-holomorphic terms in \eqref{eq:Omega-het} can also be confronted
with \eqref{eq:F1-nonholo} and one finds agreement (again, modulo a
factor $4\pi$) provided $n=2(k+2)+3$ (here we have included the four
gauge fields associated with the extra $N=2$ gravitino multiplets) and
$\chi=48$. However, this seems a numerical coincidence and we stress
that \eqref{eq:F1-nonholo} is strictly speaking only applicable to
$N=2$ supersymmetric models.

As an application we can now give the expressions of the measure for
the mixed partition function as it appears in
\eqref{eq:partition1-osv}.  Because the mixed partition function
usually refers to the holomorphic part of $\mathcal{F}_\mathrm{E}$, we
extract the non-holomorphic contribution from \eqref{eq:osv2-nonholo}
and absorb it into measure, so that the factor $\sqrt{\Delta^-}$ is
replaced by $\sqrt{\Delta^-}
\,\exp[4\pi\,\Omega^{(1)\mathrm{nonholo}}]$.  Evaluating the
expression based on \eqref{eq:F1-nonholo}, we find the following
universal result,
\begin{equation}
  \label{eq:measure-N=2}
  \sqrt{\Delta^-}\,\mathrm{e}^{4\pi\,\Omega^{(1)\mathrm{nonholo}}} \propto
     \left[\frac{K^{(0)} }{\vert
  Y^0\vert^2}\right]^{\chi/24 -1}  \,,
\end{equation}
where we only kept the leading terms which scale with zero weight in
the large-charge limit, and we dropped an irrelevant proportionality
constant. This result applies to $N=2$ only. For the CHL models one
can perform the same calculation, employing an $N=2$ description.
Provided that one chooses $n=2(k+2)+3$, accounting again for the extra
four gauge fields belonging to the $N=2$ gravitino multiplets, one
obtains,
\begin{equation}
  \label{eq:measure-N=2/N=4}
  \sqrt{\Delta^-}\,\mathrm{e}^{4\pi\,\Omega^{(1)\mathrm{nonholo}}} \propto
    \left[\frac{K^{(0)} }{\vert Y^0\vert^2}\right]  \,. 
\end{equation}
This latter result has been confirmed for the CHL models
\cite{Shih:2005he,LopesCardoso:2006bg} based on the corresponding
microscopic degeneracy formulae
\cite{Dijkgraaf:1996it,Shih:2005uc,David:2006yn,Jatkar:2005bh}.
Observe that for the FHSV and STU models, $\chi=0$, so that the
semiclassical measure factors for these models and for the CHL models
are inversely proportional. In contrast with the $N=4$ models the
semiclassical prediction for the $N=2$ and $N=8$ models does not agree
with other results in the literature. The $N=2$ results of
\cite{Denef:2007vg} for compact Calabi-Yau manifolds are qualitatively
different as they apply to large topological string coupling, whereas
the semiclassical results refer to small coupling. Hence these two
results apply to different regimes. Actually the measure factor of
\cite{Denef:2007vg} will diverge when uniformly taking the charges and
the $Y^I$ large, which reflects the so-called entropy enigma. We
expect the semiclassical results to apply to singlecentered solutions,
which are insensitive to the entropy enigma. For the $N=8$ result of
\cite{Shih:2005he} the situation is rather different, because here the
measure factor is subleading as compared to semiclassical arguments.
This seems to indicate that the semiclassical contribution will
actually vanish in this particular case, presumably as the result of
the high degree of symmetry of the $N=8$ model.

We evaluate $\Omega^{(2)}$ for the FHSV model in the next section.
Obviously these results will only be determined up to invariant
functions, just as the non-holomorphic anomaly equation of the
topological string enables the determination of the genus-$g$
partition functions up to holomorphic terms. We will demonstrate that
the results for $\Omega^{(2)}$ do not coincide with the corresponding
expressions found for the topological string in \cite{Grimm:2007tm}.

%%%%%%%%%%%%%%%%%%%%%%%%%%%%%%%%%%%%%%%%%%%%%%%%%%%%%%%%%%%%%%%%%%%
\section{Non-holomorphic corrections and the topological string}
\label{sec:non-holom-corr}
\setcounter{equation}{0}
%%%%%%%%%%%%%%%%%%%%%%%%%%%%%%%%%%%%%%%%%%%%%%%%%%%%%%%%%%%%%%%%%%%
In this section we solve the constraints posed on $\Omega^{(2)}$ for
the FHSV model, and compare the result with that for the genus-2
partition function of the topological string. As we already indicated
previously, the two results do not agree. Obviously the discrepancy
raises a variety of questions. First of all, it is important to
realize that the transformations depend on $\Upsilon$, so that we are
dealing with an iteration in $\Upsilon$, both in the function $\Omega$
as well as in the transformation rules. This situation is crucially
different from the setting in which the non-holomorphic terms arise
for the topological string, and this explains why the two results are
different. As is well known, $\Omega^{(2)}$ encodes certain terms in
the full effective action that are not necessarily local, which arise
upon integrating out the massless modes. These terms affect the
holomorphicity that underlies the Wilsonian effective action. The full
effective Lagrangian must reproduce the physically relevant
invariances, and for that the presence of the non-holomorphic
corrections can be crucial.  Indeed we will demonstrate that the free
energy \eqref{eq:free-energy-phase} is invariant up to second order in
$\Upsilon$ in the presence of the non-holomorphic corrections. This
will be discussed for the FHSV model in subsection
\ref{sec:about-Omega-2}.

A second, even more subtle, issue concerns the electric/magnetic
duality transformations. Electric/magnetic duality is defined at the
level of the effective action and its consequences are not a priori
restricted to the Wilsonian action. This duality is not necessarily a
statement about invariances, but about equivalence classes: the same
physics can be described in the context of different electric/magnetic
duality frames with different corresponding Lagrangians. These
equivalence classes are well understood for $N=2$ supersymmetric
theories at the level of the Wilsonian action, based exclusively on
the holomorphic contributions. It is reasonable to expect that the
full effective action that includes the effect of the non-holomorphic
terms remains subjected to electric/magnetic duality, so that the
functions in terms of which the full effective Lagrangian can be
encoded, should still fall into similar equivalence classes. This
requires that one can establish the existence of a different function
encoding a different Lagrangian which is related to the former by an
electric/magnetic duality transformation induced by symplectic
rotations of the period vector. In subsection
\ref{sec:non-holom-deform} we show that this situation is indeed
realized in certain cases. We prove that upon electric/magnetic
duality, there are indeed equivalence classes of functions.
Furthermore, for the class of functions that we consider in this paper,
the free energy transforms as a function under duality. In this way
the results of subsection \ref{sec:about-Omega-2} can be understood in
a more general context.

%%%%%%%%%%%%%%%%%%%%%%%%%%%%%%%%%%%%%%%%%%%%%%%%%%%%%%%%%%%%%%%%%%%
\subsection{Duality constraints on 
$\Omega^{(2)}(S,{\bar S},T,{\bar T}, Y^0, \bar Y^0)$}
\label{sec:about-Omega-2}
%%%%%%%%%%%%%%%%%%%%%%%%%%%%%%%%%%%%%%%%%%%%%%%%%%%%%%%%%%%%%%%%%%%
In this subsection we consider the duality constraints at second order
in $\Upsilon$ and $\bar\Upsilon$. We concentrate on the FHSV model,
but the corresponding result for the STU model can be derived along
the same lines. For the CHL models the $\Omega^{(g)}$ vanish for $g>1$
so that \eqref{eq:Omega-het} represents the complete result.

We start by solving the constraints imposed by S-duality, which are
given in \eqref{eq:S-invariance}. Because the $\partial\Omega/\partial
T^a$ are S-duality invariant, we can solve the second and third
equation and write $\partial\Omega/\partial S$ and
$Y^0\partial\Omega/\partial Y^0$ in terms of two functions
transforming homogeneously under S-duality. To this end we employ the
holomorphic function $G_2(2S)=\tfrac12\partial_S\ln\eta^2(2S)$, which
transforms under S-duality as,
\begin{equation}
  \label{eq:G_2}
  G_2(2S)\to \Delta_\mathrm{S}^2 \,G_2(2S)+\tfrac12 
  \mathrm{i}c\,\Delta_\mathrm{S}\,. 
\end{equation}
Observe that to $G_2(2S)$ one can always add a modular form of weight
two but this ambiguity will be absorbed in the various functions that
we will introduce shortly. We stress that we cannot assume
holomorphicity for these functions in view of the non-holomorphic
corrections noted previously. The choice for the argument $2S$ in
\eqref{eq:G_2} is made in view of the S-duality transformations which
constitute the group $\Gamma(2)$.  We now solve the third equation
\eqref{eq:S-invariance} by writing,
\begin{equation}
  \label{eq:ydy}
Y^0 \, \frac{\partial \Omega}{\partial Y^0} = w^{(0)} +
\frac{2\,G_2(2S)}{(Y^0)^2}\,
\frac{\partial \Omega}{\partial T^a} \, \eta^{ab}
\frac{\partial \Omega}{\partial T^b} \;,
\end{equation}
where $w^{(0)}$ is invariant under S-duality. Substituting this result
into the second equation \eqref{eq:S-invariance}, we obtain the
following expression for $\partial\Omega/\partial{S}$,
\begin{equation}
  \label{eq:omega-ds}
  \frac{\partial \Omega}{\partial S} = 
  w^{(2)} - 2\,G_2(2S) \, w^{(0)} - \frac{2\,[G_2(2S)]^2}{(Y^0)^2}\, 
  \frac{\partial \Omega}{\partial T^a} \, \eta^{ab} 
  \frac{\partial \Omega}{\partial T^b} \;,
\end{equation}
where $w^{(2)}$ is now a function transforming under S-duality as
$w^{(2)}\to \Delta_\mathrm{S}^2 w^{(2)}$.

The above two equations should be integrated to yield a solution for
$\Omega$. In order to do so we first note the identity
\begin{equation}
  \label{eq:G2-G4}
  [G_2(2S)]^2 = \tfrac12 \frac{\partial G_2(2S)}{\partial S} + G_4(2S)\,,
\end{equation}
where $G_4$ is a modular form of weight four, which is proportional to
the corresponding Eisenstein function $G_4(S) = (\pi/6)^2 E_4(S)$.
This identity enables one to write the square of $G_2$ in the last
term of \eqref{eq:omega-ds} as an $S$-derivative of $G_2$, because the
term proportional to $G_4$ transforms under S-duality in such a way
that it can be absorbed into the function $w^{(2)}$.  Furthermore the
second term proportional to $w^{(0)}$ can also be related to an
$S$-derivative, as can be seen by writing it as a power series in
$Y^0$,
\begin{equation}
  \label{eq:w-0-exp}
  w^{(0)}(S,\bar S,T,\bar T,Y^0,\bar Y^0) = \sum_{m\not=0}
  \frac{ v^{m}(S,\bar S,T,\bar T, \bar Y^0)}{(Y^0)^m} \,,
\end{equation}
where the functions $v^m$ transform under S-duality as modular
forms,
\begin{equation}
  \label{eq:m-mn-S-dual}
  v^m \to \Delta_\mathrm{S}^m \, v^m  \,. 
\end{equation}
The reason that the contribution with $m=0$ is not included, is
related to the fact that such a term can not show up in \eqref{eq:ydy}
in the context of a power expansion in $Y^0$.  Using the definition of
the covariant holomorphic derivative $D_S v^m= (\partial_S-2mG_2(2S))
v^m$, we can write $2\,G_2(2S) \,w^{(0)}$ as
\begin{equation}
  \label{eq:G2-w0}
  2\,G_2(2S) \,w^{(0)}=  \sum_{m\not=0} \frac{(\partial_S-D_S)v^m}
  {m\,(Y^0)^m}  \,. 
\end{equation}
The terms proportional to $D_S v^m$ transform under S-duality exactly
as $w^{(2)}$, and can thus be absorbed into it. Hence we are left
with,
\begin{equation}
  \label{eq:omega-ds-2}
  \frac{\partial \Omega}{\partial S} = 
  w^{(2)} - \sum_{m\not=0} \frac{1} {m\,(Y^0)^m} \frac{\partial
  v^m}{\partial S}  
  - \frac1{(Y^0)^2}\, \frac{\partial
  G_2(2S)}{\partial S}\,  
  \frac{\partial \Omega}{\partial T^a} \, \eta^{ab} 
  \frac{\partial \Omega}{\partial T^b} \;. 
\end{equation}
The two equations \eqref{eq:ydy} and \eqref{eq:omega-ds-2} can be
integrated  provided the following condition holds,
\begin{equation}
  \label{eq:integrability}
  \frac{\partial \Omega}{\partial T^a} \, \eta^{ab}
  \left[4\, \frac{G_2(2S)}{Y^0} \frac{\partial^2 \Omega}{\partial S
  \partial T^b} +2\, \frac{\partial G_2(2S)}{\partial S} \frac{\partial^2
  \Omega}{\partial Y^0\partial T^b} \right] = (Y^0)^2\, \frac{\partial
  w^{(2)}}{\partial Y^0} \;.
\end{equation}

Now we concentrate on the terms $\Omega^{(g)}$ with $g=1,2$, which
depend at most quadratically on $\Upsilon$ and/or $\bar\Upsilon$. In
that case the $T$-derivatives of $\Omega$ in the above formulae can be
restricted to the corresponding derivatives of $\Omega^{(1)}$ and thus
follow from the results of the previous subsection. In particular,
these $T$-derivatives depend only on $T^a$ and $\bar T^a$. According
to \eqref{eq:integrability} it then follows that $w^{(2)}$ does not
depend on $Y^0$.

We are thus left with the first equation \eqref{eq:S-invariance},
which implies that the $T$-derivatives of $\Omega$ are S-duality
invariant. Since the derivative of $\Omega^{(1)}$ was invariant under
the first term in the S-duality variation of the $T^a$ specified in
\eqref{eq:ST-S-full}, this equation leads to, 
\begin{eqnarray}
  \label{eq:T-der-Omega-S-dual}
   \bigg(\frac{\partial\Omega^{(2)}}{\partial
      T^a}\bigg)^\prime_\mathrm{S} +
  \frac{\mathrm{i}c}{\Delta_\mathrm{S} (Y^0)^2}
  \frac{\partial^2\Omega^{(1)}}{\partial T^a\partial T^b}
   \eta^{bc}
  \frac{\partial\Omega^{(1)}}{\partial T^c} &&\nonumber\\
    - \frac{\mathrm{i}c}{\bar\Delta_\mathrm{S} (\bar Y^0)^2}
  \frac{\partial^2\Omega^{(1)}}{\partial T^a\partial \bar T^b}
   \eta^{bc}
  \frac{\partial\bar\Omega^{(1)}}{\partial \bar T^c} &=&
   \frac{\partial\Omega^{(2)}}{\partial
      T^a} \,.
\end{eqnarray}
Note that, in the approximation that we are working, the S-duality
transformation on the left-hand side will not involve any variations
of the $T^a$ as those would be of even higher order in $\Upsilon$ or
$\bar\Upsilon$. Furthermore we make use of the fact that $\Omega^{(1)}$
is real, so that we extract an overall $T^a$-derivative and
establish that, 
\begin{equation}
  \label{eq:Omega-2-dT}
  \Omega^{(2)}(S,{\bar S},T,{\bar T},Y^0,\bar Y^0) 
= - \frac{G_2(2S)}{(Y^0)^2} \,
   \frac{\partial\Omega^{(1)}}{\partial T^a} \eta^{ab}
  \frac{\partial\Omega^{(1)}}{\partial T^b}
  -  \frac{G_2(2\bar S)}{(\bar Y^0)^2} \,
   \frac{\partial\Omega^{(1)}}{\partial\bar T^a} \eta^{ab}
  \frac{\partial\Omega^{(1)}}{\bar \partial\bar T^b} + u^{(0)} \,,
\end{equation}
where $u^{(0)}$ is an S-duality invariant function quadratic in
$\Upsilon,\bar\Upsilon$. Its $S$-derivative must obviously coincide
with the first two terms on the right-hand side of
\eqref{eq:omega-ds-2} as far as they are of the same order in
$\Upsilon,\bar\Upsilon$. 

Further constraints follow from imposing the T-duality equations
\eqref{eq:T-invariance}, where we will now deal exclusively with
contributions of second order in $\Upsilon,\bar\Upsilon$. We first
consider the third equation of \eqref{eq:T-invariance} and note that
the term proportional to $Y^0 \partial \Omega/\partial Y^0$ on the
right hand side of the third equation can be dropped in this order.
Using that $\partial \Omega/ \partial S$ is invariant under T-duality
we find that third equation is solved by
\begin{equation}
Y^0 \frac{\partial \Omega^{(2)}}{\partial Y^0} = r^{(0)} + \frac{1}{2 (Y^0)^2}
\, \frac{\partial \log \Phi (T)}{\partial T^a} \, \eta^{ab} \, \frac{\partial
\Omega^{(1)}}{\partial T^b} \, \frac{\partial \Omega^{(1)}}{\partial S} \;,
\label{eq:solderzOmT}
\end{equation}
where $\frac14 \partial_T \log \Phi (T)$ acts as a connection for
T-duality, as discussed below \eqref{eq:t-connection}.  Here $r^{(0)}$
denotes a T-duality invariant function.  The first equation of
\eqref{eq:T-invariance}, on the other hand, results in 
\begin{equation}
  \label{eq:dS-Omega2-T}
  \bigg(\frac{\partial \Omega^{(2)}}{\partial S}\bigg)^\prime_{\rm T}
  + \frac{2}{\Delta_{\rm T} (Y^0)^2} 
  \frac{\partial^2 \Omega^{(1)}}{\partial S^2}
  \, T^a \, \frac{\partial\Omega^{(1)}}{\partial T^a}
  + \frac{2}{{\bar\Delta}_{\rm T} ({\bar Y}^0)^2}
  \frac{\partial^2 \Omega^{(1)}}{\partial S \partial {\bar S}}
  \, {\bar T}^a \, \frac{\partial \Omega^{(1)}}{\partial {\bar T}^a}
  = \frac{\partial\Omega^{(2)}}{ \partial S}  \;,
\end{equation}
where we used again that $\Omega^{(1)}$ is real. Following the same
steps as before, this equation is solved by
\begin{eqnarray}
  \label{eq:derSomT}
  \frac{\partial \Omega^{(2)}}{ \partial S} &=&
  s^{(0)} - \frac{1}{4(Y^0)^2}
  \, \frac{\partial^2 \Omega^{(1)}}{\partial S^2} \,
  \frac{\partial\ln\Phi (T)}{\partial T^a}\,\eta^{ab} \, 
  \frac{\partial\Omega^{(1)}}{\partial T^b} \nonumber\\
  &&{}
  - \frac{1}{4 ({\bar Y}^0)^2}
  \,\frac{\partial^2 \Omega^{(1)}}{\partial S \partial {\bar S}} \,
  \frac{\partial \ln {\bar \Phi} ({\bar T})}{\partial {\bar T}^a} \,
  \eta^{ab} \, \frac{\partial
    \Omega^{(1)}}{\partial {\bar T}^b}\;,
\end{eqnarray}
where $s^{(0)}$ denotes a T-duality invariant function.  Observe that
\eqref{eq:derSomT} is consistent with the expression
\eqref{eq:omega-ds-2} for $\partial \Omega/\partial S$ following from
S-duality invariance.  Namely, the last term in \eqref{eq:omega-ds-2} is
of the type $s^{(0)}$, while the second and third term in 
\eqref{eq:derSomT} are of the type $v^2$ and $w^{(2)}$, respectively.

All results obtained so far give rise to the following expression for
$\Omega^{(2)}$, up to an S- and T-duality invariant function,
\begin{equation}
  \label{eq:Om2res}
  \Omega^{(2)} =
  - \frac{G_2 (2S)}{(Y^0)^2}\,
  \frac{\partial \Omega^{(1)}}{\partial T^a}\,\eta^{ab}\,
  \frac{\partial \Omega^{(1)}}{\partial T^b} 
  -\frac{1}{4(Y^0)^2}\,
  \frac{\partial \ln\Phi(T)}{\partial T^a}\,\eta^{ab}\, 
  \frac{\partial\Omega^{(1)}}{\partial T^b} \, \frac{\partial
  \Omega^{(1)}}{\partial S}   + {\rm c.c} \;.
\end{equation}
The reader may verify that all previous results \eqref{eq:ydy},
\eqref{eq:Omega-2-dT}, \eqref{eq:solderzOmT} and \eqref{eq:derSomT}
are reproduced. Furthermore, the result is consistent with the
assumption that $\Omega^{(2)}$ is real. 

The result \eqref{eq:Om2res} can be confronted with the manifestly
duality invariant expression,
\begin{equation}
  \label{eq:Om2cov}
  F^{(2)}(Y) \propto \frac{1}{(Y^0)^2} \,\hat G_2(2S,2\bar S)\,
  \frac{\partial \Omega^{(1)}}{\partial T^a}\,\eta^{ab}\,
  \frac{\partial \Omega^{(1)}}{\partial T^b} \;,
\end{equation}
where $\hat G_2(S,\bar S)= G_2(S) + [2\,(S+\bar S)]^{-1}$. 
Note that the right hand side of \eqref{eq:Om2cov} is non-holomorphic.
This latter
expression is the one obtained for the topological string
\cite{Grimm:2007tm}, which is clearly invariant under the lowest order
S- and T-duality transformation
by virtue of the non-holomorphic terms in $\hat G_2$ and 
$\Omega^{(1)}$.
It is clear that the real part of
\eqref{eq:Om2cov} and \eqref{eq:Om2res} are quite different.  Indeed,
$\Omega^{(2)}$ is {\it not} duality invariant in leading order of
$\Upsilon$ and $\bar\Upsilon$. It varies as follows under S- and
T-duality,
\begin{eqnarray}
  \label{eq:1ST-dauls-omega-2}
  \big(\Omega^{(2)}\big)^\prime_\mathrm{S} &=& \Omega^{(2)}
  -  \left(\frac{\mathrm{i} c}{2\,\Delta_\mathrm{S}(Y^0)^2}
  \,\frac{\partial\Omega^{(1)}}{\partial T^a}\eta^{ab} 
  \frac{\partial\Omega^{(1)}}{\partial T^b} + \mathrm{c.c.} \right)\,,
  \nonumber\\ 
  \big(\Omega^{(2)}\big)^\prime_\mathrm{T} &=& \Omega^{(2)}
  - \left(\frac{2 }{\Delta_\mathrm{T}(Y^0)^2}\, T^a
  \frac{\partial\Omega^{(1)}}{\partial T^a}\, 
  \frac{\partial\Omega^{(1)}}{\partial S} + \mathrm{c.c.} \right)\,.
\end{eqnarray}
The lack of invariance poses no problem as the function $\Omega^{(1)}$
is invariant in lowest order of $\Upsilon$ or $\bar\Upsilon$, but
still receives corrections from variations of $S$ and $T$ and their
complex conjugates that are themselves linear in $\Upsilon$ or
$\bar\Upsilon$.  This leads to the following variations, quadratic in
$\Upsilon, \bar\Upsilon$,
\begin{eqnarray}
  \label{eq:1ST-dauls-omega-1}
  \big(\Omega^{(1)}\big)^\prime_\mathrm{S} &=& \Omega^{(1)}
  + \left(\frac{\mathrm{i} c}{\Delta_\mathrm{S}(Y^0)^2}
  \,\frac{\partial\Omega^{(1)}}{\partial T^a}\eta^{ab} 
  \frac{\partial\Omega^{(1)}}{\partial T^b} + \mathrm{c.c.} \right)\,,
  \nonumber\\ 
  \big(\Omega^{(1)}\big)^\prime_\mathrm{T} &=& \Omega^{(1)}
  + \left(\frac{4}{\Delta_\mathrm{T}(Y^0)^2}\, T^a
  \frac{\partial\Omega^{(1)}}{\partial T^a}\, 
  \frac{\partial\Omega^{(1)}}{\partial S} + \mathrm{c.c.} \right)\,.
\end{eqnarray}
Observe that $\Delta_\mathrm{T}$ can be replaced by its lowest-order
value $T^a\eta_{ab}T^b$ in the second equation of
\eqref{eq:1ST-dauls-omega-2} and of \eqref{eq:1ST-dauls-omega-1}.
With these results one can verify that \eqref{eq:Om2res} also satisfies the
second equation in \eqref{eq:T-invariance}. This follows directly from
the second equations in \eqref{eq:1ST-dauls-omega-2} and
\eqref{eq:1ST-dauls-omega-1}, taking into account that all fields
$T^a$, $S$ and $Y^0$, as well as their complex conjugates, transform
under T-duality. Hence we have established that $\Omega$ satisfies the
restrictions posed by the dualities to second order in (real)
$\Upsilon$.

While $\Omega^{(1)}+ \Omega^{(2)}$ is not invariant, the quantity
$\mathrm{Im} [\Upsilon\partial_\Upsilon F]\propto [\Upsilon\partial_\Upsilon +
\bar\Upsilon\partial_{\bar\Upsilon}]\Omega$ is invariant for real
values of $\Upsilon$ at this level of approximation.  Therefore it
follows that the free energy defined in \eqref{eq:free-energy-phase}
is indeed invariant under S- and T-duality to second order in real
$\Upsilon$!

For genus $g>2$ the deviations between the functions that encode the
full effective action and the topological string twisted partition
functions will persist. The reason is that both the function $\Omega$
and the duality transformation rules depend on $\Upsilon$, which is in
striking contrast to the situation in the topological string, where
the duality transformations are independent of $\Upsilon$ and
determined, once and for all, by the classical contribution of the
function $F$. Therefore the twisted partition functions, $F^{(g)}$, of
the topological string must be different from the contributions
appearing in $\Omega$. The former are invariant under the dualities
whereas the latter are not invariant, but they are determined by the
requirement that the corresponding periods transform according to the
correct monodromy transformations. 

In section \ref{sec:introduction} we have already pointed out how this
discrepancy can possibly be resolved. The topological string partition
functions correspond to certain string amplitudes
\cite{Antoniadis:1993ze,Bershadsky:1993cx}, which are also encoded in
the full effective action that describes all the irreducible graphs.
On the other hand the latter is not invariant under duality, unlike
the partition functions of the topological string. Therefore the
information contained in the topological string and in the relevant
terms of the effective action can certainly be in agreement, although
the corresponding mathematical expressions are different. It is
suggestive that the connected and irreducible graphs are related by a
Legendre transform, whereas the action (or its underlying function)
can also be converted to an invariant expression (e.g. an Hamiltonian
or a Hesse potential) by a Legendre transform. Obviously resolving
these subtleties is a challenge.

%%%%%%%%%%%%%%%%%%%%%%%%%%%%%%%%%%%%%%%%%%%%%%%%%%%%%%%%%%%%%%%
\subsection{Non-holomorphic deformations of special geometry?}
\label{sec:non-holom-deform}
%\setcounter{equation}{0}
%%%%%%%%%%%%%%%%%%%%%%%%%%%%%%%%%%%%%%%%%%%%%%%%%%%%%%%%%%%%%%%
Motivated by the results of the preceding section we consider some of
the more conceptual issues related to the presence of non-holomorphic
corrections.  Let us consider electric/magnetic dualities on the
periods $(X^I,F_I)$, which take the form of $\mathrm{Sp}(2n)$
rotations. Here we do not assume that the $F_I$ are holomorphic
functions or sections.  Hence we have holomorphic and anti-holomorphic
coordinates $X^I$ and $\bar X^{\bar I}$, while the $F_I$ may depend on
both $X^I$ and $\bar X^I$.  To avoid ambiguous notation we will use
anti-holomorphic indices $\bar I$ wherever necessary. In this
subsection homogeneity properties do not play a role. 

Electric/magnetic dualities are defined by monodromy transformations
of the periods, defined in the usual way, 
\begin{eqnarray}
  \label{eq:em-duality}
  X^I&\to& \tilde X^I = U^I{}_JX^J + Z^{IJ}F_J\,,\nonumber\\
  F_I&\to& \tilde F_I= V_I{}^JF_J+ W_{IJ}X^J\,, 
\end{eqnarray}
where $U$, $V$, $Z$ and $W$ are the $(n+1)\times(n+1)$ submatrices
that constitute an element of $\mathrm{Sp}(2n+2,\mathbb{R})$. As a
result the relation between the old and the new fields, $X^I$ and
$\tilde X^I$, will no longer define a holomorphic map, and we note,
\begin{equation}
  \label{eq:dX-dF}
  \frac{\partial\tilde X^I}{\partial X^J} \equiv \mathcal{S}^I{}_J =
  U^I{}_J+Z^{IK}F_{KJ}\,, \qquad
  \frac{\partial\tilde X^I}{\partial \bar X^J}= Z^{IK}F_{K\bar J}\,,   
\end{equation}
where $F_{IJ}= \partial{F_I}/\partial X^J$ and $F_{I\bar J}=
\partial{F_I}/\partial\bar X^J$. Subsequently we consider the
transformation behaviour of the derivatives $F_{IJ}$ and
$F_{I\bar{J}}$ induced by electric/magnetic duality
\eqref{eq:em-duality}. Straightforward use of the chain rule yields
the relation,
  \begin{equation}
    \label{eq:dual-F_2-holo}
    F_{IJ} \to \tilde F_{IJ}=(V_I{}^L\hat F_{LK} + W_{IK})\,
      [\hat{\mathcal{S}}^{-1}]^K{}_J \,,
\end{equation}
where
\begin{eqnarray}
  \label{eq:hat-F_S}
  \hat F_{IJ}&=& F_{IJ} - F_{I\bar K}\,\bar{\mathcal{Z}}^{\bar K\bar
  L}\,\bar F_{\bar L J}\,,\nonumber\\
  \hat{\mathcal{S}}^I{}_J&=&   U^I{}_J+Z^{IK}\hat F_{KJ}\,,
  \nonumber\\
  \mathcal{Z}^{IJ}&=&  [\mathcal{S}^{-1}]^I{}_K\, Z^{KJ}\,. 
\end{eqnarray}
As was shown in \cite{deWit:2001pz}, $\mathcal{Z}^{IJ}$ is a symmetric
matrix by virtue of the fact that the duality matrix belongs to
$\mathrm{Sp}(2n+2,\mathbb{R})$. For the same reason
$[\hat{\mathcal{S}}^{-1}]^I{}_K\, Z^{KJ}$ is also symmetric in
$(I,J)$. Observe that $\mathcal{Z}^{IJ}$ satisfies the equation,
\begin{equation}
  \label{eq:partial-Z}
  \delta\mathcal{Z}^{IJ} = - \mathcal{Z}^{IK}\delta F_{KL}
  \mathcal{Z}^{LJ}\,. 
\end{equation}

Let us now assume that $F_{IJ}$ is symmetric in $I$ and $J$. This
symmetry implies that the $F_I$ can be written as the holomorphic
derivatives of some function $F(X,\bar X)$. It is of interest to
determine whether this symmetry is preserved under duality. In general
this is not the case. However, when we assume that
\begin{equation}
  \label{eq:real-mixed}
  F_{I\bar J}= \pm \bar F_{\bar JI}\,,
\end{equation}
then $\hat F_{IJ}$ will also be symmetric. In that case one can derive
from \eqref{eq:dual-F_2-holo} that $\tilde F_{IJ}$ must be symmetric
as well, so that the $\tilde F_I$ can be expressed as the holomorphic
derivatives of some function $\tilde F(\tilde X,\tilde{\bar X})$ with
respect to $\tilde X^I$. This is a first indication that
non-holomorphic deformations satisfying \eqref{eq:real-mixed} can be
consistent with the special geometry transformations of the
periods. Henceforth we will assume that \eqref{eq:real-mixed}
holds. Observe that terms in $F$ that depend exclusively on $\bar
X^{\bar I}$ are not determined by the above arguments.

Furthermore one can show that
\begin{equation}
  \label{eq:dual-F_2-mixed}
    F_{I\bar J}\to\tilde F_{I\bar
      J}=[\hat{\mathcal{S}}^{-1}]^K{}_I\,
    [\bar{\mathcal{S}}^{-1}]^{\bar L}{}_{\bar J}\,F_{K\bar L} \,. 
\end{equation}
It seems that the holomorphic and anti-holomorphic indices are treated
somewhat asymmetrically in this transformation rule. However, noting
the relation
\begin{equation}
  \label{eq:hat-S-S-1}
  (\mathcal{S}^{-1} \hat{\mathcal{S}})^I{}_J =  \delta^I{}_J -
  \mathcal{Z}^{IK} F_{K\bar L} \bar{\mathcal{Z}}^{\bar L\bar M} \bar
  F_{\bar M J}\,,
\end{equation}
which follows from \eqref{eq:hat-F_S}, and upon inverting the above
expression and writing it as a power series, one observes that
$\mathcal{S}^K{}_I \,\bar{\mathcal{S}}^{\bar L}{}_{\bar J} \,\tilde
F_{K\bar L}$ takes a more symmetric form. This enables one to show
that \eqref{eq:dual-F_2-mixed} can be expressed in two ways,
\begin{equation}
  \label{eq:dual-F_2-mixed-alt}
    F_{I\bar J}\to\tilde F_{I\bar J}
    =[\hat{\mathcal{S}}^{-1}]^K{}_I\,
    [\bar{\mathcal{S}}^{-1}]^{\bar L}{}_{\bar J}\,F_{K\bar L}=
    [{\mathcal{S}}^{-1}]^K{}_I\, 
    [\bar{\hat{\mathcal{S}}}^{-1}]^{\bar L}{}_{\bar J}\,F_{K\bar L}
    \,.  
\end{equation}

Let us now assume that the function $F$ depends on some auxiliary real
parameter $\eta$ and consider partial derivatives with respect to it.
A little calculation shows that $\partial_\eta F_I$ transforms in the
following way,
\begin{equation}
  \label{eq:auxiliary-eta}
  \partial_\eta \tilde F_I = [\hat S{}^{-1}]^J{}_I \left[
  \partial_\eta F_J - F_{J\bar K} \,\bar{\mathcal{Z}}{}^{\bar K\bar L}
  \,\partial_\eta \bar F_{\bar L}\right] \,,
\end{equation}
where the $\eta$-derivative in $\partial_\eta \tilde F_I(\tilde
X,\tilde{\bar X};\eta)$ is a partial derivative that does not act on the
arguments $\tilde X^I$ and their complex conjugates, and likewise, in
$\partial_\eta F_I(X,\bar X;\eta)$ the arguments $X^I$ and their complex
conjugates are kept fixed. Let us now assume that the function
$F(X,\bar X;\eta)$ decomposes into a holomorphic function of $X^I$ and
a purely imaginary function that depends on $X^I$, its complex
conjugates, and on the auxiliary parameter $\eta$,
\begin{equation}
  \label{eq:special-eta-function}
  F(X,\bar X;\eta)= F^{(0)}(X) + 2\mathrm{i}\,\Omega(X,\bar X;\eta)\,,
\end{equation}
where $\Omega$ is real, just as the functions we have been considering
in this paper. For this class of functions we have the following
identities, 
\begin{equation}
  \label{eq:reality}
  F_{I\bar J}= - \bar F_{\bar JI},\,, \qquad \partial_\eta F_{\bar I}=
  - \partial_\eta \bar F_{\bar I} \,,
\end{equation}
so that we must adopt the minus sign in \eqref{eq:real-mixed}. With
this result we can establish that
\begin{equation}
  \label{eq:3}
  \partial_\eta\tilde F(\tilde X, \tilde{\bar X};\eta) = \partial_\eta
  F(X,\bar X;\eta)\,,
\end{equation}
up to terms that no longer depend on $X^I$ and $\bar X^{\bar I}$.
Ignoring such terms on the ground that they are not relevant for the
vector multiplet Lagrangian, this implies that the first derivative of the
function $F$ with respect to some auxiliary parameter transforms as a
function under electric/magnetic duality. Of course, it is crucial
that we assumed the decomposition \eqref{eq:special-eta-function} so that 
$\eta$ appears only in the non-holomorphic component
$\Omega$ of $F$.

When the electric/magnetic duality defines a symmetry, then it follows
that $\partial_\eta F$ must be invariant under this symmetry. 
As we explained previously, 
S- and T-duality requires real values of $\Upsilon$.
The above arguments 
can now be applied to the free energy for BPS black holes
defined in \eqref{eq:free-energy-phase}, with the real 
$\Upsilon$ playing the role of the auxiliary parameter
$\eta$. Therefore the second term in the free energy proportional to
the $\Upsilon$-derivative of $F$ is duality invariant while the first
term equals the symplectic product of the period vector and its
complex conjugate. As a result the free energy is thus duality
invariant.

We stress once more that the effective action encoded in a
non-holomorphic function $F$ is not fully known. Although the arguments
presented above indicate that, indeed, non-holomorphic deformations
are possible within the context of special gometry, a lot of
work remains to be done in order to establish the full consistency and
the implications of this approach.

%%%%%%%%%%%%%%%%%%%%%%%%%%%%%%%%%%%%%%%%%%%%%%%%%%%%%%%%%%%%%%%%%%%%
\section{The STU model}
\label{sec:STU}
\setcounter{equation}{0}
%%%%%%%%%%%%%%%%%%%%%%%%%%%%%%%%%%%%%%%%%%%%%%%%%%%%%%%%%%%%%%%%%%%%
The analysis of the last section can be repeated for the STU model,
and undoubtedly the results will be rather similar. Nevertheless, we
still turn to a detailed analysis of this model to confront our
general results with the proposal of \cite{David:2007tq} for the
statistical degeneracies in the STU model. The STU model is based on
four fields, $Y^0$, $Y^1$, $Y^2$ and $Y^3$, of which the latter three
appear symmetrically. The fields $S$, $T$, and $U$ are defined by
$S=-\mathrm{i} Y^1/Y^0$, $T=- \mathrm{i} Y^2/Y^0$ and $U=-\mathrm{i}
Y^3/Y^0$. Much of the information has already been given in section
\ref{sec:omega-duality-constraints}. The T-duality group is contained
in $\mathrm{SO}(2,2)\cong \mathrm{SL}(2)\times\mathrm{SL}(2)$, and the
combined S- and T-duality group is the product group
$\Gamma(2)_S\times \Gamma(2)_T\times \Gamma(2)_U$, where
$\Gamma(2)\subset \mathrm{SL}(2;\mathbb{Z})$ with $a, d\in
2\,\mathbb{Z} +1$ and $b,c\in 2\,\mathbb{Z}$, with $ad-bd=1$.
Furthermore there exists a triality symmetry according to which one
can interchange $Y^1,Y^2,Y^3$ or, equivalently $S,T,U$. Under this
interchange the corresponding $\Gamma(2)$ factors of the duality
groups are interchanged accordingly.

The distinction between S- and T-duality disappears for this model and
in view of that the set-up adopted in section
\ref{sec:omega-duality-constraints} is not the most convenient one. However, we
can simply start from the S-duality as explained there and recover
the other $\Gamma(2)$ factors upon interchanging the corresponding
moduli. Hence we start from \eqref{eq:S-duality}, which we present on
the corresponding charges, 
\begin{equation}
    \label{eq:S-charge-duality}
    \begin{array}{rcl}
      p^0 &\to& d \, p^0 + c\, p^1 \;,\\
      p^1 &\to& a \, p^1 + b \, p^0 \;,\\
      p^2 &\to& d\, p^2 - c \,q_3  \;,\\
      p^3 &\to& d\, p^3 - c \,q_2  \;,
    \end{array}
    \quad
    \begin{array}{rcl}
      q_0 &\to& a\,  q_0 -b\,q_1 \;, \\
      q_1 &\to& d \,q_1 -c\, q_0 \;,\\
      q_2 &\to& a \, q_2 -b\, p^3 \;,\\
      q_3 &\to& a\, q_3 - b\, p^2  \;.
    \end{array}
\end{equation}
The T-duality (U-duality) transformations are now obtained upon
interchanging the labels $1\leftrightarrow2$ ($1\leftrightarrow3$).
From these transformation rules it follows that the eight charges
transform according to the $(\mathbf{2},\mathbf{2},\mathbf{2})$
representation of $\Gamma(2)_S\times\Gamma(2)_T\times\Gamma(2)_U$.
Consequently the charge bilinears transform as $\Gamma(2)$ triplets,
$(\mathbf{3},\mathbf{1},\mathbf{1})+
(\mathbf{1},\mathbf{3},\mathbf{1})+(\mathbf{1},\mathbf{1},\mathbf{3})$,
or in the $(\mathbf{3},\mathbf{3},\mathbf{3})$ representation. Only
the triplets are relevant for what follows and we start by defining
the following three charge bilinears,
\begin{eqnarray}
  \label{eq:charge-bilinears}
  \langle Q,Q\rangle_s &=& 2\,(q_0p^1 -q_2q_3)\,, \nonumber\\
  \langle P,P\rangle_s &=& - 2\,(q_1p^0 + p^2p^3)\,,  \nonumber\\
  \langle P,Q\rangle_s&=& q_0 p^0 -q_1p^1 +q_2p^2+q_3p^3 \,,
\end{eqnarray}
which are invariant under $\Gamma(2)_T\times\Gamma(2)_U$ and transform
as a vector under $\Gamma(2)_S$,
\begin{eqnarray}
  \label{eq:S-triplet}
  \langle Q,Q\rangle_s &\to& a^2\, \langle Q,Q\rangle_s + b^2\, \langle
  P,P\rangle_s + 2\,ab \,\langle P,Q\rangle_s \,, \nonumber\\
  \langle{P,P}\rangle_s &\to& c^2 \,\langle Q,Q\rangle_s  + d^2 \,\langle
  P,P\rangle_s  + 2\,cd \,\langle P,Q\rangle_s  \,,  \nonumber\\
  \langle P,Q\rangle_s&\to& ac\,\langle Q,Q\rangle_s + bd\, \langle
  P,P\rangle_s+ (ad+bc) \,\langle P,Q\rangle_s \,. 
\end{eqnarray}
The $\Gamma(2)_S$ invariant norm of this vector, 
\begin{equation}
  \label{eq:def-D}
  D(p,q) \equiv \langle{Q,Q}\rangle_s\,\langle{P,P}\rangle_s-\langle
  P,Q\rangle_s^{\,2} \,,
\end{equation}
is also invariant under triality, so that the two triplets of charge
bilinears that follow from \eqref{eq:charge-bilinears} by triality,
have the same invariant norm. These two other triplets, $(\langle
Q,Q\rangle_t,\langle P,P\rangle_t,\langle P,Q\rangle_t)$ and $(\langle
Q,Q\rangle_u,\langle P,P\rangle_u,\langle P,Q\rangle_u)$, transform as a
vector under $\Gamma(2)_T$ and $\Gamma_U(2)$, respectively, and are
singlets under the two remaining $\Gamma(2)$ subgroups.

In the next subsections we discuss the macroscopic determination of
the entropy of large and small black holes based on the entropy
function \eqref{eq:Sigma-simple} and the free energy
\eqref{eq:free-energy-phase}, which will include the non-holomorphic
corrections. Subsequently we consider the statistical degeneracy
formula for the STU model proposed in \cite{David:2007tq}.
 
%%%%%%%%%%%%%%%%%%%%%%%%%%%%%%%%%%%%%%%%%%%%%%%%%%%%%%%%%%%%%%%
\subsection{Macroscopic evaluation of the BPS entropy}
\label{sec:macr-eval-entr}
%%%%%%%%%%%%%%%%%%%%%%%%%%%%%%%%%%%%%%%%%%%%%%%%%%%%%%%%%%%%%%%

Here we apply the results of the preceding sections and determine
the attractor equations and the black hole entropy including the first
non-trivial subleading corrections.  For convenience we recall the
relations \eqref{eq:F_I} for the STU model,
\begin{eqnarray}
  \label{eq:F_I-STU}
  F_0&=& \frac{Y^1Y^2Y^3}{(Y^0)^2} -\frac{2\mathrm{i}}{Y^0}
  \left[-Y^0 \frac{\partial}{\partial Y^0}+
  S\frac{\partial}{\partial S}+ T\frac{\partial}{\partial T} +
  U\frac{\partial}{\partial U} 
  \right]\Omega \;,  \nonumber\\ 
  F_1&=&{}- \frac{Y^2Y^3}{Y^0} +\frac{2}{Y^0} \,
  \frac{\partial\Omega}{\partial S} \;,\nonumber\\
  F_2&=& {}- \frac{Y^1Y^3}{Y^0} +\frac{2}{Y^0} \,
  \frac{\partial\Omega}{\partial T} \;, \nonumber\\
  F_3&=& {}- \frac{Y^1Y^2}{Y^0} +\frac{2}{Y^0} \,
  \frac{\partial\Omega}{\partial U} \;,
\end{eqnarray}
which clearly exhibits the triality symmetry, provided that $\Omega$
is triality invariant. Under $\Gamma(2)_S$ the fields transform as
follows (c.f.  \eqref{eq:full-S}),
\begin{equation}
  \label{eq:full-S-STU}
  \begin{array}{rcl}
  Y^0&\to& \Delta_{\mathrm{S}}\, Y^0\;,\\[.4ex]
  Y^2&\to& \Delta_{\mathrm{S}}\, Y^2 - \displaystyle{\frac{2\, c}{Y^0} \,
  \frac{\partial\Omega}{\partial U}}  \;, 
  \end{array}
  \qquad 
  \begin{array}{rcl}
  Y^1&\to& a\,Y^1+ b\,Y^0\;, \\[.4ex]
  Y^3&\to& \Delta_{\mathrm{S}}\, Y^3 - \displaystyle{{\frac{2\,c}{Y^0} \,
  \frac{\partial\Omega}{\partial T}}}  \;.  
  \end{array}
\end{equation}
This result leads to the following transformations of the special
coordinates (c.f. \eqref{eq:ST-S-full}), 
\begin{equation}
  \label{eq:STU-full}
  S \rightarrow
  \frac{a\,S-\mathrm{i}b}{\mathrm{i}c\,S+d} \;, 
  \qquad T\to T +\frac{2\mathrm{i} c}{\Delta_{\mathrm{S}}\,(Y^0)^2}
  \,\frac{\partial\Omega}{\partial U}  \;,
\qquad U\to U +\frac{2\mathrm{i} c}{\Delta_{\mathrm{S}}\,(Y^0)^2}
  \,\frac{\partial\Omega}{\partial T}  \;.
\end{equation}

Requiring these transformations to induce the corresponding
variations on the periods, we obtain (c.f. \eqref{eq:S-invariance}), 
\begin{eqnarray}
  \label{eq:S-invariance-STU}
  \left(\frac{\partial\Omega}{\partial T}\right)^\prime_\mathrm{S} &=&
  \frac{\partial\Omega}{\partial T} \;,\qquad\quad
  \left(\frac{\partial\Omega}{\partial U}\right)^\prime_\mathrm{S} ~=~
  \frac{\partial\Omega}{\partial U} \;,
 \nonumber\\
  \left(\frac{\partial\Omega}{\partial S}\right)^\prime_\mathrm{S} -
  \Delta_{\mathrm{S}}{}^2\,\frac{\partial\Omega}{\partial S} &=&
  \frac{\partial(\Delta_{\mathrm{S}}{}^2)}{\partial{S}} 
  \left[-\tfrac12 Y^0 \frac{\partial\Omega}{\partial Y^0}
  -\frac{\mathrm{i}c}{\Delta_{\mathrm{S}}\,(Y^0)^2}
  \,\frac{\partial\Omega}{\partial T} 
  \frac{\partial\Omega}{\partial U} \right] \;,
  \nonumber\\ 
  \left(Y^0\frac{\partial\Omega}{\partial Y^0}\right)^\prime_\mathrm{S} &=&
  Y^0 \frac{\partial\Omega}{\partial Y^0} 
  +\frac{4\mathrm{i}c}{\Delta_{\mathrm{S}}\,(Y^0)^2}
  \,\frac{\partial\Omega}{\partial T}
  \frac{\partial\Omega}{\partial U}\;.
\end{eqnarray}
Corresponding results under T- and U-duality follow directly
by triality.  Subsequently we evaluate the free energy,
\begin{eqnarray}
  \label{eq:free-energy-STU}
  \mathcal{F} &=& - \vert Y^0\vert^2 (S+\bar S)(T+\bar T)(U+\bar U)
  +4\, \Omega^{(1)} 
  \nonumber\\
  &&{}
  -2\left\{\frac{\bar Y^0}{Y^0}\Big[(S+\bar S)
    \frac{\partial\Omega^{(1)}}{\partial S} 
    +(T+\bar T) \frac{\partial\Omega^{(1)}}{\partial T}
    +(U+\bar U)\frac{\partial\Omega^{(1)}}{\partial U}\Big] 
    + \mathrm{h.c.}\right\}  \,,
\end{eqnarray}
where we dropped all the higher-order $\Upsilon$ contributions.
Henceforth we will consistently restrict $\Omega$ to $\Omega^{(1)}$,
but we will nevertheless keep writing $\Omega$ for notational clarity.
The above free energy is invariant under S-, T- and U-duality, up to
terms that are quadratic in $\Omega^{(1)}$, as can be verified by
explicit calculation. These higher-order terms will eventually be
cancelled by variations of the higher-order $\Omega^{(g)}$.

Expressing $Y^2$ and $Y^3$ in terms of the charges and the field $S$,
\begin{eqnarray}
  \label{eq:Y2+3}
  Y^2 &=& \frac1{S+\bar S}\left\{ -q_3 +\mathrm{i}\bar S\,p^2
  -2\mathrm{i}\left( \frac{\partial_U\Omega}{Y^0}
  -\frac{\partial_{\bar U}\Omega}{\bar Y^0} \right)\right\} \,,
  \nonumber   \\
  Y^3 &=& \frac1{S+\bar S}\left\{ -q_2 +\mathrm{i}\bar S\,p^3
  -2\mathrm{i}\left( \frac{\partial_T\Omega}{Y^0}
  -\frac{\partial_{\bar T}\Omega}{\bar Y^0} \right)\right\} \,,
\end{eqnarray}
and imposing the remaining magnetic attractor equations, $Y^1-\bar
Y^1=\mathrm{i} \,p^1$ and $Y^0-\bar Y^0=\mathrm{i} \,p^0$, one finds,
\begin{eqnarray}
  \label{eq:nonholoSigma}
  \Sigma(S,\bar S,p,q) =
  - \,\frac{\langle{Q,Q}\rangle_s - \mathrm{i} \langle{P,Q}\rangle_s\,
  (S - {\bar S}) + \langle{P,P}\rangle_s \,|S|^2} 
    {S + {\bar S}}  + 4\, \Omega(S,\bar S,T,\bar T, U,\bar U)\;,  
\end{eqnarray}
where $T$ and $U$ are no longer independent variables but denote the
$S$-dependent values of the moduli that follow from \eqref{eq:Y2+3} to
first order in $\Omega$. To evaluate those we use the definitions,
\begin{eqnarray}
  \label{eq:P-Q}
  \begin{array}{rcl} 
    Q(S) &=& q_0+\mathrm{i}Sq_1 \,,\\
  P(S)&=& p^1-\mathrm{i}Sp^0\,,
   \end{array} 
   \qquad
   \begin{array}{rcl}
    Q_2(S)&=& q_2+ \mathrm{i}S\, p^3 \,,\\
    Q_3(S)&=& q_3+ \mathrm{i}S\, p^2 \,.
    \end{array}
\end{eqnarray}
transforming under S-duality as $P(S)\to \Delta^{-1}_S \,P(S)$, and
likewise for $Q(S)$, $Q_2(S)$ and $Q_3(S)$. Furthermore we note the
expression
\begin{equation}
  \label{eq:Y0}
    Y^0= \frac{\bar P(\bar S)}{S+\bar S} \;,
\end{equation}
so that \eqref{eq:Y2+3} leads to the following $S$-dependent
expressions for $T$ and $U$,
\begin{eqnarray}
  \label{eq:T-U}
  T&=&\mathrm{i}\, \frac{\bar Q_3(\bar S)}{\bar P(\bar S)} -
  \frac{2\,(S+\bar S)}{\bar P(\bar S)} 
  \left(\frac{\partial_U\Omega}{\bar P(\bar S)}
  -\frac{\partial_{\bar U}\Omega}{P(S)} \right)\,,
  \nonumber\\
  U&=& \mathrm{i}\,\frac{\bar Q_2(\bar S)}{\bar P(\bar S)} -
  \frac{2\,(S+\bar S)}{\bar P(\bar S)} 
  \left(\frac{\partial_T\Omega}{\bar P(\bar S)}
  -\frac{\partial_{\bar T}\Omega}{P(S)} \right)\,.
\end{eqnarray}
Observe that the S-duality transformation of these equations coincides
with the results \eqref{eq:STU-full}. For what follows we need to
evaluate the derivatives of $\bar T$ and $\bar U$ with respect to $S$,
\begin{eqnarray}
  \label{eq:partialT-barS}
  \frac{\partial\bar T}{\partial S}&=& -\tfrac12 \langle{P,P}\rangle_u\,
  P^{-2}(S)+\cdots\,, \nonumber\\
  \frac{\partial\bar U}{\partial S}&=& -\tfrac12 \langle{P,P}\rangle_t\,
  P^{-2}(S)+\cdots\,,
\end{eqnarray}
where we suppressed terms proportional to the derivatives of $\Omega$.

Finally the attractor equation for $S$ follows from requiring the
$S$-derivative of \eqref{eq:nonholoSigma} to vanish, 
\begin{eqnarray}
  \label{eq:S-attractor}
  &&{}
  \langle{Q,Q}\rangle_s + 2\mathrm{i}\,\langle{P,Q}\rangle_s \bar S
  -\langle{P,P}\rangle_s \,\bar S^2 \nonumber \\
  &&{}
  +2(S+\bar S)^2 \left\{2\,\partial_S\Omega -
  \frac{\langle{P,P}\rangle_u}{P^2(S)} 
  \,\partial_{\bar T}\Omega - \frac{\langle{P,P}\rangle_t}{P^2(S)}
  \,\partial_{\bar U}\Omega\right\} = 0 \,. 
\end{eqnarray}
It is important to check the behaviour of this result under the
various dualities. It is covariant under S-duality, because, in this
approximation, the term proportional to the derivatives of $\Omega$
scale under S-duality with the same factor
$\bar\Delta_\mathrm{S}^{-2}$ as the other terms in
\eqref{eq:S-attractor}.

In the following, we will consider large black holes, i.e. black holes
with charges such that $D(p,q)>0$, and hence with $\langle
P,P\rangle_s\not= 0$. In that case the solution of
\eqref{eq:S-attractor} takes the following form,
\begin{eqnarray}
  \label{eq:S-attractor-sol}
   S&=& \sqrt{\frac{D}{\langle{P,P}\rangle_s^{\,2}}}\left\{1+
  \frac4{\langle{P,P}\rangle_s}\left[2\,\partial_{\bar S}\Omega -
  \frac{\langle{P,P}\rangle_u}{\bar P^2(\bar S)} 
  \,\partial_{T}\Omega - \frac{\langle{P,P}\rangle_t}{\bar P^2(\bar S)}
  \,\partial_{U}\Omega\right] \right\} \nonumber \\
   &&{}
  -\frac{\mathrm{i}\,\langle{P,Q}\rangle_s}{\langle{P,P}\rangle_s} \,,
\end{eqnarray}
where the arguments in $\Omega$ are the leading values of $S,\bar
T,\bar U$ as our results hold only to first order of in $\Omega$. At
this point it is easy to substitute these values for $S$ into
\eqref{eq:T-U} and we find the same equations for the fixed-point
values for $T$ and $U$ as in \eqref{eq:S-attractor-sol} upon triality
transformations. These results are the extension of the lowest-order
expressions that were obtained long ago \cite{LopesCardoso:1996yk}. 

Before considering the behaviour under T- and U-duality of
\eqref{eq:S-attractor-sol}, we note the following identities, which
hold at the attractor point, 
\begin{eqnarray}
  \label{eq:dual-QP}
  2\, \langle{P,P}\rangle_s\,\vert P(S)\vert^2 &=& -
  \langle{P,P}\rangle_t\,\langle{P,P}\rangle_u +\cdots
  \,,\nonumber\\[.4ex] 
  2\, \langle{P,P}\rangle_s\, Q_2(S)\,\bar P(\bar S) &=& 
  \langle{P,P}\rangle_t\,\langle{P,Q}\rangle_u  -\tfrac12\mathrm{i}\,
  \langle{P,P}\rangle_s\langle{P,P}\rangle_t \, (S+\bar S)
  + \cdots\,, \nonumber\\
  T+\bar{T} &=& -\tfrac12  \langle{P,P}\rangle_u \,\frac{S+\bar
  S}{\vert P(S)\vert^2} + \cdots \,, 
\end{eqnarray}
as well as similar identities obtained by triality. Furthermore we
note the transformations,
\begin{equation}
  \label{eq:P(S)-transf}
  P(S)\buildrel S\over\longrightarrow \frac{P(S)}{\Delta_S}\,,\qquad
  P(S) \buildrel {\mathrm{T,U}}\over\longrightarrow
  \bar\Delta_{\mathrm{T,U}} \,P(S) + \cdots \,. 
\end{equation}
With these equations one establishes that the expression
\eqref{eq:S-attractor-sol} for $S$ transforms under T- and U-duality as,
\begin{equation}
  \label{eq:TU-variation-S}
  S\buildrel{\mathrm{T,U}}\over\longrightarrow S+ \frac{2\mathrm{i}
  c_{T,U}}{\Delta_{\mathrm{T,U}}\,(Y^0)^2} 
  \,\partial_{U,T}\Omega   \;,
\end{equation}
which is precisely compatible with \eqref{eq:STU-full} upon triality.

Now we can introduce a modified field $S^\mathrm{inv}$
invariant under T- and U-duality by 
\begin{eqnarray}
  \label{eq:S-inv-attractor-sol}
   S^\mathrm{inv}&=& \sqrt{\frac{D}{\langle{P,P}\rangle_s^{\,2}}}\left\{1+ 
  \frac{8\,\partial_{\bar{S}}\Omega} {\langle{P,P}\rangle_s}\right\}  
  -\frac{\mathrm{i}\,\langle{P,Q}\rangle_s}{\langle{P,P}\rangle_s} \,,
\end{eqnarray}
which transforms in the usual way under S-duality as it is the
solution of an S-duality covariant equation,
\begin{eqnarray}
  \label{eq:S-inv-attractor}
  &&{}
  \langle{Q,Q}\rangle_s +
  2\mathrm{i}\,\langle{P,Q}\rangle_s \bar S
  -\langle{P,P}\rangle_s \,\bar S^2 + 4(S+\bar S)^2 \,\partial_S\Omega
  = 0 \,. 
\end{eqnarray}
This equation results from the condition that (we set $\Upsilon=-64$)
\begin{eqnarray}
  \label{eq:nonholoSigma-eff}
  \Sigma^\mathrm{S}(S^\mathrm{inv},\bar S^\mathrm{inv};p,q) &=& 
  - \,\frac{\langle{Q,Q}\rangle_s - \mathrm{i} \langle{P,Q}\rangle_s\,
  (S^\mathrm{inv} - {\bar S}^\mathrm{inv}) + \langle{P,P}\rangle_s
  \,|S^\mathrm{inv}|^2} 
    {S^\mathrm{inv} + \bar S^\mathrm{inv}}  \nonumber\\ 
    &&{}
    -\frac{2}{\pi}\,\ln\big[\vert\vartheta_2(S^\mathrm{inv})\vert^4
  (S^\mathrm{inv}+\bar S^\mathrm{inv}) \big] \,,
\end{eqnarray}
is stationary. Likewise we can introduce similar equations for fields
$T^\mathrm{inv}$ and $U^\mathrm{inv}$ which transform as usual under
T- and U-duality respectively, but are invariant under the other
dualities. These fields are the solutions of the equations that follow
from \eqref{eq:S-inv-attractor} by triality.

The result for the entropy now follows from substituting the value of
$S$ into \eqref{eq:nonholoSigma}. All the $\Omega$-dependent terms in
the solutions for $S,T,U$ cancel generically, and one is left with
\eqref{eq:nonholoSigma} with $S$ (and thus $T$ and $U$ in $\Omega$)
equal to their classical values. Observe that this is so because we
are only considering the first-order corrections to the entropy. In
principle there are higher-order terms which will represent
next-to-subleading corrections to the entropy. The result for the
entropy thus takes the form,
\begin{eqnarray}
  \label{eq:STU-entropy}
  \mathcal{S}_\mathrm{STU}(p,q)&=&
  \pi\,\Sigma\Big\vert_\mathrm{attractor} \nonumber\\[.5ex] 
  &=& {}
  \pi\,\sqrt{D(p,q)} - 2\,\ln\big[\vert\vartheta_2(S)\vert^4\,(S+\bar
  S)\big] 
  \nonumber\\ 
  &&{} 
  -2\,\ln\big[\vert\vartheta_2(T)\vert^4\,(T+\bar T)\big]
  -2\,\ln\big[\vert\vartheta_2(U)\vert^4\,(U+\bar U)\big] \,, 
\end{eqnarray}
where, in the last terms $S$, $T$ and $U$ are fixed to their
lowest-order attractor values. Here we made use of
\eqref{eq:Omega-STU}. 

Alternatively, 
the entropy \eqref{eq:STU-entropy} can be obtained from an
entropy function $\tilde\Sigma$ that depends on the invariant fields
$S^\mathrm{inv}$, $T^\mathrm{inv}$ and $U^\mathrm{inv}$, where these
fields are treated as independent. This entropy function is given by
\begin{equation}
  \label{eq:inv-entropy-function}
  \tilde\Sigma(S^\mathrm{inv},\bar S^\mathrm{inv},T^\mathrm{inv},\bar
  T^\mathrm{inv},U^\mathrm{inv},\bar U^\mathrm{inv};p,q) = \tfrac13
  \left[\tilde\Sigma^\mathrm{S} +\tilde\Sigma^\mathrm{T}
  +\tilde\Sigma^\mathrm{U} \right] \,,
\end{equation}
where $\tilde\Sigma^\mathrm{S}$ is S-, T- and U-duality invariant and
equal to,
\begin{eqnarray}
  \label{eq:nonholoSigma3}
  \tilde\Sigma^\mathrm{S}(S^\mathrm{inv},\bar S^\mathrm{inv};p,q) &=& 
  - \,\frac{\langle{Q,Q}\rangle_s - \mathrm{i} \langle{P,Q}\rangle_s\,
  (S^\mathrm{inv} - {\bar S}^\mathrm{inv}) + \langle{P,P}\rangle_s
  \,|S^\mathrm{inv}|^2} 
    {S^\mathrm{inv} + \bar S^\mathrm{inv}}  \nonumber\\ 
    &&{}
    -\frac{6}{\pi}\,\ln\big[\vert\vartheta_2(S^\mathrm{inv})\vert^4
  (S^\mathrm{inv}+\bar S^\mathrm{inv}) \big] \,,
\end{eqnarray}
and $\tilde\Sigma^\mathrm{T}$ and $\tilde\Sigma^\mathrm{U}$ follow by
triality. Extremizing $\tilde\Sigma$ with respect to $S^\mathrm{inv}$,
$T^\mathrm{inv}$ and $U^\mathrm{inv}$ and substituting the resulting
values into $\tilde\Sigma$ yields the entropy \eqref{eq:STU-entropy},
where we work in the same order of approximation as before. Note,
however, that $\tilde\Sigma^\mathrm{S}$ does not equal
\eqref{eq:nonholoSigma-eff} so that the value of the attractor point
will be different, although, at this order of approximation, such a
deviation has no effect on the entropy. 

%%%%%%%%%%%%%%%%%%%%%%%%%%%%%%%%%%%%%%%%%%%%%%%%%%%%%%%%%%%%
\subsection{Small black holes} 
%%%%%%%%%%%%%%%%%%%%%%%%%%%%%%%%%%%%%%%%%%%%%%%%%%%%%%%%%%%%
To explore some other aspects of the STU model, we now consider
possible small black hole solutions.  Small black holes satisfy
$D(p,q) = 0$, with $D$ given in \eqref{eq:def-D}.  The
higher-curvature corrections encoded in $\Omega$ are then crucial to
ensure that the moduli are attracted to finite values at the horizon.
For the STU model, the associated $\Omega^{(1)}$, given in
\eqref{eq:Omega-STU}, depends on all three moduli $S, T$ and $U$,
which implies that in order for the three moduli to take finite values
at the horizon, the charges carried by the small black hole have to be
chosen in such a way as to result in three non-vanishing charge
bilinears (out of the nine bilinears introduced earlier).  This
differs from the situation encountered in $N=4$ models, where the
associated $\Omega^{(1)}$ only depends on one modulus, so that only
one non-vanishing charge bilinear is required to construct a small
black hole \cite{Dabholkar:2004yr}.

An obvious possibility consists in choosing charges such that only
$\langle{Q,Q}\rangle_s$, $\langle{Q,Q}\rangle_t$ and
$\langle{Q,Q}\rangle_u$ are different from zero. Such a configuration
can be obtained by switching on the charges $q_0,q_1,q_2,q_3$ while
leaving the remaining ones equal to zero, so that
$\langle{Q,Q}\rangle_s= -2\,q_2q_3$, $\langle{Q,Q}\rangle_t= - 2\, q_1
q_3$ and $\langle{Q,Q}\rangle_u= -2\, q_1q_2$. Then, at the horizon,
$Y^0,Y^1,Y^2,Y^3$
are all real, so that $S,T,U$ are purely imaginary, which does not
constitute a well-behaved situation (since, for instance, 
the non-holomorphic
terms contained in $\Omega^{(1)}$ are expressed in terms of 
the real part of the moduli fields).
Therefore, we discard this
choice of charges and take instead $p^0,q_2,q_3$ as non-vanishing
charges.  Then, the non-vanishing charge bilinears are,
\begin{equation}
  \label{eq:bilinears-em-small-bh}
  \langle{Q,Q}\rangle_s= -2\,q_2q_3\,,\quad  \langle{P,P}\rangle_t= - 2\,
  p^0 q_2 \,,\quad \langle{P,P}\rangle_u= - 2\, p^0q_3 \,. 
\end{equation}
In that case $Y^1,Y^2,Y^3$ are real, 
but $Y^0$ is not in view of the
fact that $p^0\not=0$.  Using the definition of $S, T$ and $U$ we establish
the following expressions for these
quantities, 
\begin{eqnarray}
  \label{eq:Y-em-small-bh}
  \begin{array}{rcl}
    Y^0 &=&   \mathrm{i} \bar S\,{\displaystyle\frac {p^0}{S+\bar S}}
  \,,\\[4mm] 
  Y^1 &=&   - \bar S S\,{\displaystyle\frac{ p^0}{S+\bar S}} \,,
  \end{array} 
  \qquad
  \begin{array}{rcl}
      Y^2 &=&   - \bar S T\,{\displaystyle\frac{ p^0}{S+\bar S}}
      \,,\\[4mm]
      Y^3 &=&   - \bar S U \,{\displaystyle\frac{p^0}{S+\bar S}} \,,
  \end{array}
\end{eqnarray}
so that $\bar S U$ and $\bar S T$ are real.  Inserting 
\eqref{eq:Y-em-small-bh} into \eqref{eq:F_I-STU}
and restricting $\Omega$ to $\Omega^{(1)}$ gives,
\begin{eqnarray}
  \label{eq:F-small-bh}
  F_0 &=& \frac{1}{\bar S^2} \left\{\frac{p^0\,STU\bar S^3}{S+\bar S} -
  \frac{2(S+\bar S)\bar S}{p^0} \big[S\partial_S\Omega+ T
  \partial_T\Omega+U\partial_U\Omega\big] \right\}\,, \nonumber\\  
  F_1 &=& \frac{\mathrm{i}}{\bar S} \left\{\frac{p^0\,TU\bar
  S^2}{S+\bar S}   -
  \frac{2(S+\bar S)}{p^0} \,\partial_S\Omega\right\}\,, \nonumber\\ 
  F_2 &=& \frac{\mathrm{i}}{\bar S} \left\{\frac{p^0\,US\bar
  S^2}{S+\bar S}   -
  \frac{2(S+\bar S)}{p^0} \,\partial_T\Omega\right\}\,, \nonumber\\ 
  F_3 &=& \frac{\mathrm{i}}{\bar S} \left\{\frac{p^0\,TS\bar S^2}{S+\bar S}-
  \frac{2(S+\bar S)}{p^0} \,\partial_U\Omega  \right\} \,.
\end{eqnarray}
Using ${\bar T} = {\bar S} T/S$ and ${\bar U} = {\bar S} U/S$, 
we find that the attractor equations $F_0 = {\bar F}_{\bar 0}$ and 
$F_1 = {\bar F}_{\bar 1} $
yield, respectively,
\begin{eqnarray}
( S - {\bar S}) 
{\bar S}^2 T U &=& \frac{2}{(p^0)^2}  
(S + {\bar S}) \left( S^2 \partial_S \Omega
- {\bar S}^2 \partial_{\bar S} \Omega \right.
\nonumber\\
&& \qquad \qquad \qquad 
\left.
+  S T \partial_T \Omega - 
{\bar S} {\bar T} \partial_{\bar T} \Omega 
+  S U \partial_U \Omega - 
{\bar S} {\bar U} \partial_{\bar U} \Omega
\right) \;, \nonumber\\
{\bar S}^2 T U &=& \frac{2}{(p^0)^2}  
(S + {\bar S}) \left( S \partial_S \Omega
+ {\bar S} \partial_{\bar S} \Omega \right) \;,
\label{eq:F1-eq}
\end{eqnarray}
while the attractor equations 
$F_2 - {\bar F}_{\bar 2}= i q_2 $ and
$F_3 - {\bar F}_{\bar 3} = i q_3$ read, 
\begin{eqnarray}
{\bar S} U &=& \frac{q_2}{p^0} 
+ \frac{2}{(p^0)^2} 
 \frac{(S + \bar S)}{|S|^2 } \left( S \partial_T \Omega + 
{\bar S} \partial_{\bar T} \Omega \right) \;, \nonumber\\
{\bar S} T &=& \frac{q_3}{p^0} 
+ \frac{2}{(p^0)^2} 
\frac{(S + \bar S)}{|S|^2 } \left( S \partial_U \Omega + 
{\bar S} \partial_{\bar U} \Omega \right) \;.
\label{eq:F2-eq}
\end{eqnarray}
In the absence of higher-curvature corrections, inspection of \eqref{eq:F1-eq}
and \eqref{eq:F2-eq} shows that there are no solutions with finite values
of $S, T$ and $U$. 
When including higher-curvature corrections, on the other hand, we 
deduce from the structure of \eqref{eq:F1-eq} and \eqref{eq:F2-eq} 
that a likely solution exists with
finite, but small values for $T$ and $U$, and 
a large, but finite value for $S$.  Therefore, we  
expand $\Omega$ around large values of $S$ and small values of $T$ and $U$.
Using $\Omega^{(1)}$ given in \eqref{eq:Omega-STU}, 
we obtain accordingly (with $\Upsilon = - 64$),
\begin{equation}
\Omega_{\rm STU}^{(1)} = \frac14 \, (S + \bar S) 
- \frac {1}{2 \,\pi} \left(
\log (S + \bar S) + 
\log (\frac{1}{T} + \frac{1}{\bar T}) + \log 
(\frac{1}{U} + \frac{1}{\bar U})   \right) \;.
\label{eq:OmegaSTUlarge}
\end{equation}
Here we used \eqref{eq:eta-function} in the expansion of $\vartheta_2$.
Observe that the non-holomorphic terms in $\Omega^{(1)}$ are crucial
for obtaining finite horizon values for $T$ and $U$.

Using \eqref{eq:OmegaSTUlarge} we find that the first equation in
\eqref{eq:F1-eq} is identical to the second equation in
\eqref{eq:F1-eq} multiplied by $S - {\bar S}$.  This means that $S -
\bar S$ does not get determined at the horizon.
The second equation yields
\begin{equation}
{\bar S}^2 T U = \frac{2}{(p^0)^2}  
(S + {\bar S}) \left( \frac14 (S + \bar S) - \frac{1}{2 \pi} \right) \;,
\label{eq:valueSpbS}
\end{equation}
while from \eqref{eq:F2-eq} we obtain
\begin{eqnarray}
{\bar S}U &=& 
\frac{q_2}{p^0} + \frac{S + \bar S}{\pi \, (p^0)^2 \, 
{\bar S} T} \;,  \nonumber
\\
{\bar S} T &=& \frac{q_3} {p^0} + \frac{S + \bar S}{\pi \, (p^0)^2 \, 
{\bar S} U} \;.
\label{eq:values STSU}
\end{eqnarray}
Thus we see that the attractor equations determine the values of 
$S + {\bar S}, {\bar S} U$ and ${\bar S} T$, while the remaining
moduli are left undetermined.

In the following, we take $p^0, q_2, q_3$ to be positive and uniformly
large.  For large $S + {\bar S}$, \eqref{eq:valueSpbS} can be
approximated by $S + {\bar S} = \sqrt{2} p^0 \sqrt{{\bar S}^2 T U}$,
while \eqref{eq:values STSU} implies that ${\bar S}U$ and ${\bar S} T$
are of order one with approximate values given by
\begin{equation}
{\bar S}U = \frac{|\langle Q,Q\rangle_s|}{|\langle P,P\rangle_u|}
\;\;\;,\;\;\;
{\bar S} T = \frac{|\langle Q,Q\rangle_s|}{|\langle P,P\rangle_t|}
 \;,
\label{eq:SUSTbilin}
\end{equation}
where we made use of the charge bilinears 
\eqref{eq:bilinears-em-small-bh}.
Reinserting this into $S + {\bar S}$ gives
\begin{equation}
S + {\bar S}= \sqrt { |\langle Q,Q\rangle_s|} \;.
\label{eq:SSbilin}
\end{equation}

The entropy of this small black hole can be computed using
\eqref{eq:nonholoSigma} at the attractor point.  Its value is entirely
determined in terms of \eqref{eq:SUSTbilin} and \eqref{eq:SSbilin}.
We obtain, up to an additive constant,
\begin{equation}
{\cal S_{\rm macro}} = 2 \pi \, \sqrt { |\langle Q,Q\rangle_s|}
- 2 \log \left( \frac{|\langle P,P\rangle_t \, \langle P,P\rangle_u|}{
\sqrt { |\langle Q,Q\rangle_s|}} \right) \;.
\label{eq:SmacroSTUsmall}
\end{equation}
We note that for large charges, the leading term in the entropy 
depends only on one of the bilinears \eqref{eq:bilinears-em-small-bh}.
This is in contrast to what one naively obtains when 
considering the microstate degeneracy proposal of \cite{David:2007tq} and
evaluating the degeneracy integral on an electric or magnetic divisor.
There one expects to obtain a microscopic degeneracy which,
to leading order, is given by the sum of three terms, each involving
the square root of one of the three charge bilinears 
\eqref{eq:bilinears-em-small-bh}.  This, however, is in conflict with 
\eqref{eq:SmacroSTUsmall}, which indicates the need for a better
understanding of the microstate degeneracy proposal of 
\cite{David:2007tq}.

%%%%%%%%%%%%%%%%%%%%%%%%%%%%%%%%%%%%%%%%%%%%%%%%%%%%%%%%%%%%%%%%%%%%
\subsection{Comparison with microstate degeneracies}
\label{sec:comp-micro-deg}
%\setcounter{equation}{0}
%%%%%%%%%%%%%%%%%%%%%%%%%%%%%%%%%%%%%%%%%%%%%%%%%%%%%%%%%%%%%%%%%%%%
Recently, a proposal \cite{David:2007tq} was put forward for the
microscopic degeneracies of twisted sector dyons in the STU model in
terms of the residues of certain products of Siegel modular forms, and
it was shown that the leading and subleading results for the entropy
of these dyons agree with the macroscopic analysis that we have
presented in subsection \ref{sec:macr-eval-entr}.  Here we briefly
review the analysis of the asymptotic degeneracies based on the
microscopic formula in the notation of \cite{LopesCardoso:2006bg}. It
is based on the procedure used earlier in
\cite{Cardoso:2004xf,Jatkar:2005bh}. The degeneracy of dyons depends
on the residues of the inverse of a modular form $\Phi_0 (\rho,
\sigma,\upsilon)$ of weight zero under a subgroup of
$\mathrm{Sp}(2;\mathbb{Z})$. The three modular parameters,
$\rho,\sigma,\upsilon$, parametrize the period matrix of an auxiliary
genus-two Riemann surface which takes the form of a complex,
symmetric, two-by-two matrix. For the STU model the proposed
degeneracies are given by the product of three of the following
integrals over appropriate 3-cycles,
%which take the following form,
\begin{eqnarray}
   \label{eq:dphi0}
   I(K,L,M) \propto \oint \mathrm{d}
 \rho\,\mathrm{d}\sigma\,\mathrm{d}\upsilon \;  
 \frac{{\rm e}^{\mathrm{i} \pi [
     \rho\, K + \sigma\, L + (2 \upsilon -1)\, M]}}
 {\Phi_0(\rho,\sigma,\upsilon)} \;.
\end{eqnarray}
The quantities $K,L,M$ are integers proportional to the charge
bilinears $\langle{P,P}\rangle$, $\langle{Q,Q}\rangle$ and
$\langle{P,Q}\rangle$, and thus transform as triplets under
$\Gamma(2)$. The inverse of the modular form $\Phi_0$ takes the form
of an infinite Fourier sum with integer powers of $\exp[\pi\mathrm{i}
\rho]$, $\exp[\pi\mathrm{i}\sigma]$ and $\exp[2\pi\mathrm{i}
\upsilon]$, and the 3-cycle is then defined by choosing integration
contours where the real parts of $\rho$ and $\sigma$ take values in
the interval $(0,2)$ and the real part of $\upsilon$ takes values in
the interval $(0,1)$.  The leading behaviour of the dyonic degeneracy
is associated with the rational quadratic divisor ${\cal D} = \upsilon
+ \rho \sigma - \upsilon^2=0$ of $\Phi_{0}$, near which $1/\Phi_0$
takes the form,
\begin{eqnarray}
   \label{eq:Phik}
   \frac{1}{\Phi_0 (\rho,\sigma,\upsilon)}  \approx
   \frac{1}{\mathcal{D}^2} \; \frac{\sigma^{2}}
   { f^{(0)} (\gamma') \, f^{(0)}(\sigma')} 
   +\mathcal{O}(\mathcal{D}^0) \;,
\end{eqnarray}
where
\begin{eqnarray}
   \label{eq:gsprime}
   \gamma' = \frac{ \rho \sigma - \upsilon^2}{\sigma} \;,\qquad 
 \sigma' = \frac{\rho \sigma - (\upsilon-1)^2}{\sigma} \;,
\end{eqnarray} 
and $f^{(0)}(\gamma)= \vartheta_2^{\,4}(\gamma)$. The divisor is
invariant under the following $\Gamma(2)$ transformations, 
\begin{eqnarray}
\label{eq:modparatrans}
    \rho &\to& a^2\,\rho +b^2\,\sigma -2\,ab\,\upsilon +ab\,, \nonumber\\
  \sigma &\to& c^2 \rho + d^2 \,\sigma -2\,cd\,\upsilon+cd\,, \nonumber\\  
  \upsilon &\to& {}-ac\, \rho -bd\,\sigma +(ad+bc) \upsilon-bc \,,
\end{eqnarray}
which belong to the invariance group of $\Phi_0$. With this
information it can be verified straightforwardly that the
function~\eqref{eq:dphi0} is therefore invariant under $\Gamma(2)$
using that $K,L,M$ transform precisely as the charge bilinears in
\eqref{eq:S-triplet}.

As stated above, 
the proposal for the dyon degeneracy reads, 
\begin{equation}
  \label{eq:deg-STU}
  d_\mathrm{STU}(p,q) = I(K_s,L_s,M_s)\;I(K_t,L_t,M_t)\;I(K_u,L_u,M_u)\;,
\end{equation}
which is manifestly invariant under triality.
When performing an asymptotic evaluation of the integral
(\ref{eq:dphi0}), one must specify which limit in the charges is
taken. Large black holes correspond to a limit where both electric and
magnetic charges are taken to be large. More precisely, one takes $K
L-M^2\gg 1$, and $K+L$ must be large and negative. 
Under a uniform scaling of the
charges the field $S^\mathrm{inv}$ given in
\eqref{eq:S-inv-attractor-sol} will then remain finite; to ensure that
it is nevertheless large one must assume that $\vert K\vert$ is
sufficiently small as compared to $\sqrt{KL -M^2}$. In this way one
can recover the non-perturbative string corrections, as was stressed
in \cite{Cardoso:2004xf}.

Clearly, $\Phi_0(\rho_s,\sigma_\sigma,\upsilon_s)$ has double zeros at
$\upsilon_{s\pm} = \ft12 \pm \ft12 \sqrt{1 +4 \rho_s \sigma_s}$ on
the divisor.  The evaluation of the integral (\ref{eq:dphi0}) proceeds
by first evaluating the contour integral for $\upsilon$ around either
one of the poles $\upsilon_{s\,\pm}$, and subsequently evaluating the two
remaining integrals over $\rho_s$ and $\sigma_s$ in saddle-point
approximation.  The saddle-point values of $\rho_s, \sigma_s$, and hence
of $\upsilon_{s\pm}$, can be parametrized by
\begin{eqnarray}
\label{eq:rsvfixed}
\rho_s = \frac{\mathrm{i} |S^\mathrm{inv}|^2}{S^\mathrm{inv} + \bar
 S^\mathrm{inv}} \;, \qquad 
 \sigma_s = \frac{\mathrm{i}}{S^\mathrm{inv} + \bar S^\mathrm{inv}}
\;,\qquad \upsilon_{s\pm} = \frac{S^\mathrm{inv}}{S^\mathrm{inv} + \bar
 S^\mathrm{inv}} \;,
\end{eqnarray}
with $S^\mathrm{inv}$ given in
\eqref{eq:S-inv-attractor-sol}.\footnote{%%%%%%%%%%%%  
  Observe that $\rho$, $\sigma$, $\upsilon$ constitute the complex
  two-by-two period matrix, which appears in the exponential factor of the
  integrand in \eqref{eq:dphi0} sandwiched between the charge vectors.
  At the divisor, the imaginary part of this matrix is proportional to
  the coset representative of $\mathrm{SO}(2,1)/\mathrm{SO}(2)$,
  parametrized by the invariant dilaton field. 
} %%%%%%%%%%%%%%%%%%%%%%%%%%%%%%%%%%%%%%%%%%%%%%%%%%%% 
The same considerations apply to the other integrals in
\eqref{eq:deg-STU} with identical results. As argued in
\cite{Cardoso:2004xf}, these values describe the unique solution to
the saddle-point equations for which the state degeneracy $d(p,q)$
takes a real value. The resulting expression for $\log d_{\rm STU}
(p,q)$ precisely equals the expression for the macroscopic entropy
(\ref{eq:STU-entropy}), with $S$ (and similarly $T$ and $U$) expressed
in terms of the charges through the first term in
(\ref{eq:S-attractor-sol}). The result is valid up to a constant and
up to terms that are suppressed by inverse powers of the charges.
Other divisors are expected to give rise to exponentially suppressed
corrections to the microscopic entropy $\mathcal{S}_{\rm micro} = \log
d_{\rm STU} (p,q)$. This result is in accordance with the generic features
of the semiclassical approximation that we have outlined in
section \ref{sec:partition-entropy-function}. 

The microstate degeneracy proposal of \cite{David:2007tq} does,
however, raise a few questions which in our mind indicate that a
better understanding of the microstate degeneracy is needed.  First of
all, the saddle-point equation for $S^\mathrm{inv}$ resulting from the
asymptotic evaluation of \eqref{eq:dphi0}, is the one following from
\eqref{eq:nonholoSigma3} and therefore it does not agree with the
attractor equation \eqref{eq:S-inv-attractor} derived from the
macroscopic analysis.  This is in contrast to the situation
encountered in the $N=4$ models discussed in
\cite{Cardoso:2004xf,Jatkar:2005bh}.

Second, when considering the small black hole discussed in
\eqref{eq:SmacroSTUsmall}, it is not clear how the microstate proposal
\eqref{eq:deg-STU} can reproduce the leading term of the entropy of this
small black hole.  In the case of a small black hole, the degeneracy
integral \eqref{eq:dphi0} needs to be evaluated on either an electric
or a magnetic divisor, and to leading order this yields a contribution
to the microscopic entropy proportional to the square root of the
appropriate charge bilinear.  Since the microscopic degeneracy
proposal \eqref{eq:deg-STU} involves three integrals, with each
integral contributing a term of this type, the resulting microscopic
entropy consists of a sum of three terms, each involving the square
root of one of the three charge bilinears
\eqref{eq:bilinears-em-small-bh}.  This, however, is in conflict with
\eqref{eq:SmacroSTUsmall}.

Finally, we have considered the computation of the mixed black hole partition
function, as was done in the context of $N=8$ \cite{Shih:2005he} and
$N=4$ \cite{Shih:2005he,LopesCardoso:2006bg} models, in the hope of reproducing
\eqref{eq:partition1-osv}. Hence we start from the definition of  the
mixed black hole partition function \eqref{eq:partition1-osv} with
$d_{\rm STU} (p,q)$ expressed by \eqref{eq:deg-STU}, and with $K,L,M$
given by the charge bilinears $\langle P,P \rangle, \langle Q,Q
\rangle$ and $\langle P,Q \rangle$ (here we omit a proportionality
factor between these two sets of bilinears, for simplicity).  The
summation over $q_0$ leads to a delta function, whereas the sum over
$q_1,q_2,q_3$ can be done by a Poisson resummation. In this way we
obtain the following result,
\begin{eqnarray}
  \label{eq:STU-mixed-partition}
  &&Z_\mathrm{STU}(p,\phi) =
  \sum_{\phi\mathrm{-shifts}}\oint\oint\oint
  \frac1{\sqrt{\sigma_s\sigma_t\sigma_u}
  \,\Phi_0(\rho_s,\sigma_s,\upsilon_s)
  \,\Phi_0(\rho_t,\sigma_t,\upsilon_t)
  \,\Phi_0(\rho_u,\sigma_u,\upsilon_u) }\\[.3ex]
  &&{}
  \times\delta\big(\phi^0 +\mathrm{i}p^0(2v_s+2v_t+2v_u -3) + 2\mathrm{i}
  (p^1\sigma_s +p^2\sigma_t +p^3\sigma_u )\big) \nonumber\\[.8ex]
  &&{}
  \times \exp\left(-2\pi\mathrm{i} \left[p^2p^3\rho_s +p^3p^1\rho_t
  +p^1p^2\rho_u  - \frac{\phi^{s2}+\phi^{t2}+\phi^{u2} 
  -2(\phi^s\phi^t+\phi^t\phi^u+\phi^u\phi^s)  }
  {16\,\sigma_s\sigma_t\sigma_u}  
  \right] \right) \,, \nonumber
\end{eqnarray}
where the sum over shifts of $\phi$ are by arbitrary integer steps of
$2\mathrm{i}$. The quantities $\phi^s$, $\phi^t$ and $\phi^u$ are
given by
\begin{eqnarray}
  \label{eq:phi-stu}
  \phi^s= \sigma_s \phi^1 -2\mathrm{i} p^0 \rho_s\sigma_s
  - \mathrm{i}p^1 \sigma_s (2\upsilon_s
  -2\upsilon_t-2\upsilon_u+1) \,,
\end{eqnarray}
with $\phi^t$ and $\phi^u$ related by triality. The resulting integral
is supposed to be a function of $\phi^0,\phi^1,\phi^2,\phi^3$, and of
the charges $p^0,p^1,p^2,p^3$, but this feature is no longer manifest
in the expression \eqref{eq:STU-mixed-partition}. Unlike in the $N=4$
models, it is a non-trivial task to explicitly evaluate the integral,
although it should, for instance, be possible to use a saddle-point
approximation and make contact with semiclassical predictions.  \\
{\bf Note added}: Meanwhile this problem has been addressed in
\cite{Cardoso:2008ej}.

%%%%%%%%%%%%%%%%%%%%%%%%%%%%%%%%%%%%%%%%%%%%%%%%%%%%%%%%%%%%%%%
\section{Conclusion}
\label{sec:conclusion}
\setcounter{equation}{0}
%%%%%%%%%%%%%%%%%%%%%%%%%%%%%%%%%%%%%%%%%%%%%%%%%%%%%%%%%%%%%%%%
In this paper we demonstrated that non-holomorphic corrections are
crucial for obtaining a BPS black hole free energy that is manifestly
invariant under duality transformations.  In our approach, these
corrections are encoded in a single real homogeneous function
$\Omega$, in order to ensure that the attractor equations will still
follow by requiring stationarity of the free energy. We presented
evidence that these corrections describe a consistent non-holomorphic
deformation of special geometry. The precise relationship between the
non-holomorphic terms encoded in $\Omega$ and the effective
supersymmetric action remains to be worked out. 

In the context of $N=2$ models with exact duality symmetries, such as
the FHSV and the STU models, an explicit evaluation of the
non-holomorphic corrections to $\Omega$ reveals that these are related
to, but quantitatively different from the non-holomorphic corrections
to the topological string.  This difference may be related to the
Legendre transformation that transforms the holomorphic prepotential
of complex special geometry into the real Hesse potential of real
special geometry.  The latter is related to the BPS black hole free
energy and therefore manifestly duality invariant. It would be very
interesting to investigate this further. 

Duality invariance of the black hole partition function also requires
the presence of a non-trivial integration measure when writing the BPS
degeneracies in the form of an inverse Laplace transform over a mixed
partition function \cite{Ooguri:2004zv}. We gave a prediction for the
measure factor for a class of $N=2$ black holes using semiclassical
arguments, which, however, disagrees with the results for string
compactifications based on compact Calabi-Yau manifolds at strong
topological string coupling \cite{Denef:2007vg}. A direct test of our
semiclassical prediction for the measure factor requires knowledge of
the exact microscopic state degeneracy. When confronting our
macroscopic results for large and small black holes in the STU model
with the microstate degeneracy proposal of \cite{David:2007tq}, we
identify a number of subtle issues that to us indicate the need for a
better understanding of the microstate degeneracy of the STU model.

%%%%%%%%%%%%%%%%%%%%%%%%%%%%%%%%%%%%%%%%%%%%%%%%%%%%%%%%%%%%%%%%%
\subsection*{Acknowledgements}
We acknowledge helpful discussions with Ignatios Antoniadis, Nathan
Berkovits, Michele Cirafici, Justin David, Jan de Boer,  Frederik
Denef, Albrecht Klemm, Marcos Mari\~no, Thomas Mohaupt, Hirosi Ooguri,
Ashoke Sen, Stephan Stieberger, and Tom Taylor. B.d.W. thanks the
\'Ecole Normale Sup\'erieure in Paris, where part of this work was
carried out, for hospitality and the Centre National de la Recherche
Scientifique (CNRS) for financial support. The work of S.M. is
supported by research grants from The Netherlands Organisation for
Scientific Research (NWO) and the Max Planck Institut f\"ur
Gravitationsphysik, Potsdam.  S.M. would like to thank Bernard de Wit
and the members of ITP, Utrecht University, as well as Hermann Nicolai
and the members of Quantum gravity group at AEI, Potsdam for the nice
hospitality during the course of this work. This work is partly
supported by EU contracts MRTN-CT-2004-005104 and MRTN-CT-2004-512194
and by NWO grant 047017015.

%%%%%%%%%%%%%%%%%%%%%%%%%%%%%%%%%%%%%%%%%%%%%%%%%%%%%%%%%%%%%%%%%
%%%%%%%%%%%%%%%%%%%%%%%%%%%%%%%%%%%%%%%%%%%%%%%%%%%%%%%%%%%%%%%%%%
%\bibliographystyle{utphys} 
%\bibliography{refs}
\providecommand{\href}[2]{#2}
\begingroup\raggedright\endgroup
\end{document}